\documentstyle[preprint,eqsecnum,aps,epsf]{revtex}
\newif\iftightenlines\tightenlinesfalse
\tightenlines\tightenlinestrue
\begin{document}
%
%%%%%%%%%%%%%%%%%%%%%%%%%%%%%%%%%%%%%%%%%%%%%%%%%%%%%%%%%%%%%%%%%%%%%%%%%%%%%%%
\def\tG{\tilde G}
\def\ETC{E_T^c}
\def\eslt{\not\!\!{E_T}}
\def\to{\rightarrow}
\def\te{\tilde e}
\def\tf{\tilde f}
\def \tlam{\tilde{\lambda}}
\def\tl{\tilde l}
\def\tb{\tilde b}
\def\tst{\tilde t}
\def\tt{\tilde t}
\def\ttau{\tilde \tau}
\def\tmu{\tilde \mu}
\def\tg{\tilde g}
\def\tga{\tilde \gamma}
\def\tnu{\tilde\nu}
\def\tell{\tilde\ell}
\def\tq{\tilde q}
\def\tw{\widetilde W}
\def\tz{\widetilde Z}
\def\Rsl{\not\!\!{R}}
%
%%%%%%%%%%%%%%%%%%%% TITLE PAGE %%%%%%%%%%%%%%%%%%%%%%%%%%%%%%%%%%%%%%%%%%%%%%
%
%\draft
\preprint{\vbox{\baselineskip=14pt%
   \rightline{FSU-HEP-990305}\break 
   \rightline{UH-511-926-99}
}}
\title{THE REACH OF TEVATRON UPGRADES
IN GAUGE-MEDIATED 
SUPERSYMMETRY BREAKING MODELS}

\author{Howard Baer$^1$, P.~G. Mercadante$^2$,
Xerxes Tata$^2$ and Yili Wang$^2$}
\address{
$^1$Department of Physics,
Florida State University,
Tallahassee, FL 32306, USA
}
\address{
$^2$Department of Physics and Astronomy,
University of Hawaii,
Honolulu, HI 96822, USA
}
%
%\date{\today}
\maketitle
\begin{abstract}

We examine signals for sparticle production at the Tevatron within the
framework of gauge mediated supersymmetry breaking models for four
different model lines, each of which leads to qualitatively different
signatures. We identify cuts to enhance the signal above Standard Model
backgrounds, and use ISAJET to evaluate the SUSY reach of experiments at
the Fermilab Main Injector and at the proposed TeV33. For the model
lines that we have examined, we find that the reach is at least as
large, and frequently larger, than in the mSUGRA framework. For two of
these model lines, we find that the ability to identify $b$-quarks and
$\tau$-leptons with high efficiency and purity is essential for the
detection of the signal.

\end{abstract}

\medskip


%%%%%%%%%%%%%%%%%% MAIN TEXT %%%%%%%%%%%%%%%%%%%%%%%%%%%%%%%%%%%%%%%%%%%%%%%

\section{Introduction}

Models where gauge interactions~\cite{early,dine} rather than gravity
serve as messengers of supersymmetry breaking have been the focus of
many recent phenomenological
analyses~\cite{dim,ambros,babu,bbct,bagger,rs,nandi,gunion,fm,sarid,bmtw,wells,rg}
of supersymmetry (SUSY). In these models, sparticle masses and decay
patterns differ from those in the extensively studied mSUGRA
model~\cite{sugrarev} which has served as the framework for many
experimental analyses of supersymmetry. Perhaps the most important
difference between the mSUGRA framework and gauge-mediated SUSY breaking
(GMSB) models with a low SUSY breaking scale is the identity of the
lightest SUSY particle (LSP). In the former case, the lightest
neutralino ($\tz_1$) is almost always the LSP, while in the GMSB framework,
the gravitino is much lighter than other sparticles. Moreover, while the
gravitino is essentially decoupled in mSUGRA scenarios, the couplings of
the Goldstino (which forms the longitudinal components of the massive
gravitino), though much smaller than gauge couplings, may nonetheless be
relevant for collider physics in that they can cause the next to 
lightest SUSY particle (NLSP) to decay into a gravitino {\it inside the
detector}. The precise decay pattern and lifetime of the NLSP depends on
its identity and on model parameters. For instance, if $\tz_1$ is the
NLSP, it would decay via $\tz_1 \to \tG + \gamma$, and if kinematically
allowed, also via $\tz_1 \to \tG + Z$, or into the various Higgs bosons
of SUSY models via $\tz_1 \to \tG + h,\ H,$ or $A$. If, on the other hand, the
NLSP is a slepton, it would decay via $\tell \to \tG + \ell$, {\it etc.}
Sparticles other than the NLSP decay only very rarely to gravitinos, so
that it is safe to neglect these decays in any analysis. Thus heavier
sparticles cascade decay as usual to the NLSP, which then decays into
the gravitino as described above.

Sparticle signatures differ from those in the mSUGRA framework for two
basic reasons. First, if the NLSP is not the lightest neutralino, the
cascade decay patterns to the NLSP are modified. Second, the NLSP (which
need not be electrically neutral) itself decays into a gravitino and
Standard Model (SM)
particles. The gravitinos escape the experimental apparatus undetected
resulting in $\eslt$ in SUSY events. In the GMSB framework, however,
neutralino NLSP decays may also result in isolated photons or $Z$ or
Higgs bosons which could provide additional handles to reduce SM backgrounds
to the SUSY signal. If the NLSP is a slepton, all SUSY events should
contain leptons of the same flavour as the slepton NLSP in addition to
$\eslt$. While it may be possible to have other candidates for the NLSP,
this does not seem to be the case in the simplest realizations of the
GMSB framework, and we will not consider this possibility any further. 

Within the minimal GMSB framework, supersymmetry breaking in a hidden
sector is communicated to the observable sector via SM gauge
interactions of messenger particles (with quantum numbers of $SU(2)$ doublet
quarks and leptons) whose mass scale is characterized by
$M$. As a result, the soft SUSY breaking masses induced
for the various sparticles
are directly proportional to the strength of their gauge
interactions. Thus, coloured squarks are heavier than sleptons, and
gluinos are heavier than electroweak gauginos. 
The observable sector sparticle masses and couplings are
determined (at the scale $M$) by the GMSB model parameter set, 
\begin{equation}
\Lambda, M, n_5, \tan\beta, sign(\mu),C_{grav}.
\label{eq:mgmsb}
\end{equation}
Of these, $\Lambda$ is the most important parameter in that it sets the
scale of sparticle masses. The model predictions for the mass parameters
at the scale $M$ are then evolved down to the sparticle mass scale via
renormalization group evolution (RGE). Radiative breaking of electroweak
symmetry determines $|\mu|$. The weak scale SUSY parameters depend only
weakly on the messenger mass scale $M$, since this primarily enters as
the scale at which the mass relations predicted by the model are assumed
to be valid. There is an additional dependence of the sparticle spectrum
on $M$ due to threshold effects\cite{martin}, but this is also weak as
long as $M/\Lambda$ is not very close to unity. Messenger quarks and
lepton, it is assumed, can be classified into complete vector
representations of $SU(5)$: the number ($n_5$) of such multiplets is
required to be $\leq 4$ for the messenger scale $M = {\cal O}(100~TeV)$
in order that the gauge couplings remain perturbative up to the grand
unification scale.  Finally, the parameter $C_{grav} \geq 1$~\cite{bmtw}
(essentially, the ratio of hidden sector to messenger sector SUSY
breaking vevs)
%is a factor~\cite{bmtw} which determines by how much the gravitino mass is
%heavier than the minimal gravitino mass in this framework. 
can be used to dial the gravitino mass beyond its minimum value.
Effectively,
$C_{grav}$ parametrizes the rate for sparticle decays into a
gravitino. This decay is most rapid when $C_{grav}=1$, while for larger
values of $C_{grav}$ the NLSP may decay with an observable decay
vertex, or may even be sufficiently long-lived to pass all the way
through the detector. In this extreme case, SUSY event topologies would
be identical to those in the mSUGRA model if $\tz_1$ is the
NLSP. However, for the case where a charged slepton is the NLSP, SUSY
events would necessarily contain a pair of penetrating tracks from the
long-lived slepton NLSP, which might be detectable at the Tevatron as
``additional (possibly slow) muons'' \cite{fm,wells}. In our analysis, we
assume that the NLSP decays promptly and fix $C_{grav}=1$; {\it i.e.} we
do not attempt to model the additional handle displaced vertices might
provide to reduce SM backgrounds.

Many of the phenomenological implications depend only weakly on the
parameter $M$.  Thus, the $\Lambda-\tan\beta$ plane provides a
convenient panorama for illustrating the diversity of phenomenological
possibilities in GMSB scenarios. This is shown in Fig.~1 for
$M=3\Lambda$ and {\it
a})~$n_5 = 1$, {\it b})~$n_5 = 2$, {\it c})~$n_5 = 3$ and {\it d})~$n_5
= 4$. We choose $\mu$ to be positive since for this choice the model
predictions~\cite{bbct,desh} are well within experimental
constraints~\cite{cleo} from the decay $b \to s \gamma$ over essentially
the whole plane.  In Region~1 in Fig.~1{\it a}-{\it d} (the boundaries
of these regions are the heavy solid lines), the lightest neutralino is
the NLSP, so that $\tz_1 \to \tG \gamma$ (and to $Z$ and Higgs bosons if
these decays are kinematically allowed). In Region 2, $m_{\ttau_1} <
m_{\tz_1}$, while other sleptons are heavier than $\tz_1$, and cascade
decays of sparticles terminate in $\ttau_1$, except immediately above
the boundary between Regions 1 and 2 where the decay $\tz_1 \to \ttau_1
\tau$ is kinematically forbidden. For parameters in Region 2, we thus
expect an excess of $\tau$ leptons in SUSY events~\cite{nandi}. In
Regions 3 and 4 in Fig.~1{\it b-d}, not only $\ttau_1$, but also $\te_1
\simeq \te_R$ and $\tmu_1 \simeq \tmu_R$, are lighter than $\tz_1$. Thus
neutralinos are effectively sources of real (dominantly right-handed)
sleptons. In Region 3, $\tell_1 \to \ell \tG$ because its decay $\tell_1
\to \ttau_1 \tau \ell$ ($\ell =e,\mu$) is kinematically forbidden. The
decay $\tmu_1 \to \nu_{\mu} \ttau_1 \nu_{\tau}$ which occurs via
suppressed muon Yukawa couplings is kinematically allowed, and may
compete with the decay to $\tmu_1 \to \mu\tG$; for $C_{grav} = 1$, we
find that this decay (which has been included in our computation), is
unimportant. In Region~4, the decays $\tell_1 \to \ttau_1
\bar{\tau} \ell$ and $\tell_1 \to \bar{\ttau}_1 \tau \ell$
are also allowed, and compete with the gravitino decay of
$\tell_1$. Frequently, the stau decays of $\tell_1$ dominate its decays
to gravitino, and then, as for Region 2, SUSY events will be
characterized by an abundance of taus in the final state. Signals from
sparticle production will clearly depend on which of these regions the
model parameters happen to lie.

The grey regions in Fig.~1 are excluded because the observed pattern of
electroweak symmetry breaking is not obtained: in the wedge in the upper
left corner, $m_{\ttau_R}^2 < 0$ while in the band on top, $m_A^2 <
0$. The non-observation of sparticle signatures in experiments at LEP
excludes other portions of the plane. Within the MSSM charginos have
been excluded if $m_{\tw_1} \leq 90-95$~GeV.  While this limit has been
obtained assuming that charginos and selectrons decay into a stable
neutralino which escapes detection, we expect that an even more striking
signature is obtained if $\tz_1$ decays via $\tz_1 \to \gamma \tG$. The
leftmost dot-dashed line in Fig.~1 is the contour $m_{\tw_1} =95$~GeV:
to its left, charginos are lighter than 95~GeV.  To assist the reader in
assessing the sparticle mass scale, we have also shown
mass contours for $m_{\tw_1} = 200$~GeV and $m_{\tw_1} = 350$~GeV.
MSSM searches for
acollinear electron pairs exclude selectrons lighter than about 90~GeV. 
This limit should certainly be valid within this framework if
$\te_1 \to \tG e$ (Region 3)
or even if it decays to $\tz_1$ that subsequently
decays to a photon (Region 1). The dotted line is the contour $m_{\te_1} =
90$~GeV. For the case where $\te_1$ mainly decays to $\ttau_1$, the
actual bound may be somewhat weaker, and closer to the MSSM stau bound
$\sim 76$~GeV. The most stringent experimental limit for the $n_5=1$ and
$n_5=2$ cases in frames {\it a}) and {\it b}) comes from the the LEP
search \cite{lep} for $\gamma\gamma+\eslt$ events from $e^+e^- \to
\tz_1\tz_1$ production. The cross section for this process depends on
the selectron mass. The ALEPH analysis \cite{lep} for $n_5=1$ results in
the lower limit, $m_{\tz_1} \geq 84$~GeV~\cite{footnote1}.
For larger values of $n_5$, the selectron to
$\tz_1$ mass ratio is smaller, so that the corresponding cross section
is even larger than in the $n_5=1$ case. Indeed the DELPHI
collaboration~\cite{lep} has obtained a preliminary bound $m_{\tz_1}
\agt 88$~GeV for $n_5=2$, for parameters in Region 1. If $m_{\ttau_1}
\leq m_{\tz_1}$, $\tz_1$s act as sources of staus and add to the signal
from direct stau pair production. The DELPHI search for acollinear tau
pairs still limits $m_{\tz_1} \leq 86-90$~GeV, and also the bounds
$m_{\ttau_1} \leq 76$~GeV, regardless of $m_{\tz_1}$. In Fig.~1, in the
horizontally hatched region $m_{\tz_1} \leq 84$~GeV, whereas in the region
with vertical hatches, $m_{\ttau_1} \leq 76$~GeV.
%These regions are
%representative of the bounds on neutralino and stau
%masses within this framework. 
Finally, the LEP experiments \cite{higgs}
have a preliminary bound of 91-95~GeV on the mass of the SM Higgs
boson. Since, for small values of $\tan\beta$ the lighter Higgs boson $h$
of the GMSB framework is frequently close to the SM Higgs boson, we also
show the regions with $m_h \leq 95$~GeV (the  diagonally hatched area in the
lower left corner) in Fig.~1. 
Furthermore, LEP analyses exclude $m_A \alt 83$~GeV when $\tan\beta$ is
large. This excludes the thin (diagonally hatched) sliver where
$\tan\beta \sim 53$.
The reader should appreciate that the various shaded regions that
we have shown are not formal experimental limits, but indicate
the reach of present experiments within the GMSB framework.
 
We see from Fig.~1 that current experiments have already probed Regions
1 and 2 if $n_5 > 2$. On the other hand, for $n_5 =1$, we have just
these two regions, while $n_5=2$, all four regions are still possible.
Since experimental signatures within the GMSB framework differ
significantly from those in the MSSM and mSUGRA models, it is of
interest to reassess the sensitivity of Tevatron experiments to signals
from sparticle production at the upcoming Run~II of the Tevatron Main
Injector (MI) as well as at the proposed luminosity upgrade (dubbed
TeV33) where an integrated luminosity $\sim 25$~$fb^{-1}$ might be
accumulated. This is the main purpose of this paper. We had begun this
program in an earlier study \cite{bbct} where we had computed cross
sections for various SUSY event topologies for models with $n_5=1$
expected at the Tevatron: in this case, the NLSP is dominantly the
hypercharge gaugino. Here, we first repeat this analysis for somewhat
different model parameters, using cuts and acceptances more appropriate
to Run II. We have also fixed a bug in the program~\cite{footnote2}
which resulted in an underestimate of the chargino pair production cross
section. We also examine cases with larger values of $n_5$ for which we
expect the phenomenology to change qualitatively from our earlier
study. Toward this end, we first examine a model line with $n_5=2$ with
$\tan\beta=15$ where $\ttau_1$ is the NLSP and significantly lighter
than other sleptons. Next, we examine a model line with $n_5=3$ where
all right handed sleptons are roughly degenerate in mass (the co-NLSP
scenario), and where $\te_1$ ($\tmu_1$) dominantly decay via $\te_1 \to
e\tG$ ($\tmu_1 \to \mu \tG$). Finally, we examine a non-minimal model
where the NLSP is dominantly a Higgsino-like neutralino. This is not
because we believe this is any more likely than the mGMSB scenarios
previously discussed, but because it leads to qualitatively different
experimental signatures. In view of the fact that the underlying
mechanism of SUSY breaking, and hence the resulting mass pattern, is
unknown it seems worthwhile to explore implications of unorthodox
scenarios, particularly when they lead to qualitative differences in the
phenomenology.

In the next Section, we describe the upgrades that we have made to
ISAJET to facilitate the simulation of the minimal GMSB framework that
we have described, as well as several of its non-mimimal extensions. In
Sec.~III we specify four different model lines and discuss strategies
for separating the SUSY signal from SM background for each of these. Our
main result is the
projection for the reach of
experiments at the MI and at TeV33. We end in Sec.~IV with a summary of
our results together with some general remarks.

\section{Simulation of Gauge Mediated SUSY Breaking Scenarios}

We use the event generator program ISAJET v 7.40 for simulating SUSY events at
the Tevatron. Since ISAJET has been described elsewhere \cite{isajet},
we will only discuss recent improvements that we have made that
facilitate the simulation of the mGMSB model specified by the parameter
set (\ref{eq:mgmsb}), and also, some of its variants. The `GMSB option'
allows one to use the parameter set (\ref{eq:mgmsb}) as an input. ISAJET
then computes sparticle masses at the messenger scale $M$, then evolves
these down to the lower scale relevant for phenomenology, and finally
calculates the `MSSM parameters' that are then used in the evaluation of
sparticle cross sections and decay widths.  The decays of neutralinos
into gravitinos, $\tz_i \to \tG \gamma$, $\tz_i \to \tG Z$ and $\tz_i
\to \tG h,H,A$ as well as (approximately) the Dalitz decay $\tz_i \to
e^+e^-\tG$ are included in ISAJET. The decays $\tell_1 \to \ell\tG$ and
$\ttau_1 \to \tau\tG$, as well as the three body decays~\cite{AMK}
$\tell_1 \to \ttau_1 \bar{\tau}\ell$ and $\tell_1 \to \bar{\ttau_1}\tau
\ell$, which are mediated by a virtual neutralino, have also been
included.  The widths for corresponding three body decays mediated by
virtual chargino exchange are suppressed by the lepton Yukawa coupling, and
are also included. These decays can be significant only for smuons, and
only when $m_{\tmu_1}-m_{\ttau_1} \alt m_{\tau}$ so that the
neutralino-mediated three body decays of $\tmu_1$ are kinematically very
suppressed or forbidden. Although for $C_{grav}=1$ we have not found this to be
important, for larger values of $C_{grav}$, the
(long-lived) smuon may dominantly decay via the chargino mediated decay
to a stau, and may alter the apparent curvature of the `smuon track' in
the detector~\cite{footnote3}.

We have also included in ISAJET the facility to simulate several
non-minimal gauge-mediated SUSY breaking models that involve additional
parameters. While these will be irrelevant to the analysis performed in
our present study, we have chosen to describe this for completeness
because it may prove useful to readers studying extensions of the
minimal class of models. 
\begin{itemize}
\item The parameter $\Rsl$ allows the user to adjust~\cite{dtw} the
ratio between the gaugino and scalar masses by scaling the former by the
factor $\Rsl$ which is equal to unity in the minimal GMSB framework. 

\item In GMSB models, additional interactions are needed to generate the
dimensional $\mu$ and $B$ parameters that are essential from
phenomenological considerations. These interactions can split the
soft SUSY breaking masses of Higgs and lepton doublets (at the messenger
scale) even though these have the same gauge quantum numbers. These
additional contributions to the squared masses of Higgs doublets that
couple to up and down type fermions, are parametrized~\cite{dtw} by
$\delta m_{H_u}^2$ and $\delta m_{H_d}^2$, respectively. These
parameters are zero in the minimal model.

\item If the hypercharge $D$-term has a non-zero expectation value $D_Y$
in the messenger sector, it will lead to additional contributions to
sfermion masses at the messenger scale which may be parametrized~\cite{dtw} as
$\delta m_{\tf}^2 =g'YD_Y$. The value of $D_Y$ (which is zero in the
minimal GMSB framework) is constrained as it can lead to an unacceptable
pattern of electroweak symmetry.

\item Finally, allowing incomplete messenger representations~\cite{martin} can
effectively result in different numbers ($n_{5_1}$, $n_{5_2}$ and
$n_{5_3}$) for each factor of the gauge group. 
\end{itemize}
ISAJET allows the user to simulate these non-minimal models using the
GMSB2 command.

To model the experimental conditions at the Tevatron, we use the toy
calorimeter simulation package ISAPLT. We simulate calorimetry covering
$-4 \leq \eta \leq 4$ with a cell size given by $\Delta\eta \times
\Delta\phi= 0.1 \times 0.087$, and take the hadronic (electromagnetic)
calorimeter resolution to be $0.7/\sqrt{E}$ ($0.15/\sqrt{E}$). Jets are
defined as hadronic clusters with $E_T > 15$~GeV within a cone of
$\Delta R= \sqrt{\Delta\eta^2+\Delta\phi^2} = 0.7$ with $|\eta_j| \leq
3.5$.
Muons and electrons with $E_T > 7$~GeV and $|\eta_{\ell}| < 2.5$
are considered to be isolated if the the scalar sum of
electromagnetic and hadronic $E_T$ (not including the lepton, of course)
in a cone with $\Delta R=0.4$ about the lepton to be smaller than
$max(2~GeV, E_T(\ell)/4)$.
Isolated leptons are also required to be
separated from one another by $\Delta R \geq 0.3$. 
We identify photons within $|\eta_{\gamma}|< 1$ if $E_T > 15$~GeV, and
consider them to be isolated if the additional $E_T$ within a cone of
$\Delta R = 0.3$ about the photon is less than 4~GeV. Tau leptons are
identified as narrow jets with just one or three charged prongs with
$p_T>2$~GeV within $10^{\circ}$ of the jet axis and no other charged
tracks in a 30$^{\circ}$ cone about this axis. The invariant mass of
these tracks is required to be $\leq m_{\tau}$ and the net charge of the
three prongs required to be $\pm 1$. QCD jets with $E_T =15 (\geq
50)$~GeV are misidentified as taus with a probability of 0.5\% (0.1\%)
with a linear interpolation in between. Finally, for SVX tagged
$b$-jets, we require a jet (satisfying the above jet criteria) within
$|\eta_j|\leq 1$ to contain a $B$-hadron with $p_T \geq 15$~GeV. The jet
is tagged as a $b$-jet with a probability of 55\%. Charm jets (light
quark or gluon jets) are mistagged as $b$-jets with a probability of 5\%
(0.2\%).

\section{The Reach of Tevatron Upgrades for Various Model Lines}

Within the GMSB framework, sparticle signatures, and hence the reach of
experimental facilities, are qualitatively dependent on the nature of the
NLSP. Here, we examine the reach of experiments at the Tevatron Main
Injector as well as that of the proposed TeV33 upgrade for four
different model lines~\cite{footnote4} where the NLSP is (A)~dominantly
a hypercharge gaugino, (B)~the stau lepton, $\ttau_1$, with other
sleptons significantly heavier than $\ttau_1$, (C) again the stau, but
$\te_1$ and $\tmu_1$ are essentially degenerate with $\ttau_1$, and
(D)~dominantly a Higgsino. We fix the messenger scale $M = 3\Lambda$,
$\mu> 0$ and $C_{grav}=1$ throughout our analysis. We use ISAJET to
compute signal cross sections, incorporating cuts and triggers to
simulate the experimental conditions at the Tevatron together with
additional cuts that serve to separate the SUSY signal from SM
backgrounds. We project the reach of future Tevatron upgrades for each
of these scenarios.

\subsection{Model Line A: The Bino NLSP Scenario}

We see from Fig.~1{\it a} that for $n_5=1$, the lightest neutralino is
the NLSP as long as $\tan\beta$ is not very large. Since the value of
$|\mu|$ computed from radiative breaking of electroweak symmetry is
rather large, the NLSP is mainly a bino. To realize the bino NLSP model
line, we fix $\tan\beta = 2.5$ which ensures that sleptons are
significantly heavier than $m_{\tz_1}$. Sparticles cascade decay to
$\tz_1$ which then mainly decays via $\tz_1 \to \gamma\tG$. Thus almost
all SUSY events contain at least two hard isolated photons. 

In Fig.~\ref{mlA:prod}{\it a} we show the mass spectrum of sparticles
that might be in the Tevatron range versus $\Lambda$, which sets the
sparticle mass scale, while in frame {\it b}) we show the cross sections
for the most important sparticle production mechanisms at the
Tevatron. We see that chargino pair production and $\tw_1\tz_2$
production dominate because squarks and gluinos are beyond the Tevatron
reach. The production of right-handed slepton pairs is suppressed
relative to chargino/neutralino production by over an order of
magnitude. Values of $\Lambda$ smaller than $\sim 70$~TeV are excluded
by the LEP search for $\gamma\gamma +\eslt$ events.  For $\Lambda \alt
80$~TeV (corresponding to $m_{\tz_1} \alt 90-100$~GeV), the two body
decay $\tw_1 \to W\tz_1$ is kinematically suppressed, and the chargino
mainly decays via $\tw_1 \to \ttau_1\nu_{\tau}$ or $\tw_1 \to qq\tz_1$;
for $\Lambda \agt 80$~TeV, the decay $\tw_1 \to W\tz_1$ dominates. The
neutralino $\tz_2$ dominantly decays via $\tz_2 \to \tz_1 h$ (for
$\Lambda \geq 90$~TeV) when this decay is not kinematically suppressed:
otherwise it decays via $\tz_2 \to \tell_1 \ell$, with roughly equal
branching fractions for all three lepton flavours. We thus expect that
$\tw_1\tw_1$ and $\tz_2\tw_1$ production will mainly lead to jetty
events (counting hadronically decaying taus as jets) possibly with
additional $e$ and $\mu$ plus photons plus $\eslt$.

We use ISAJET to classify the supersymmetric signal events primarily by
the number of isolated photons --- events with $<2$ photons arise when
one or more of the photons is outside the geometric acceptance, has too
low an $E_T$, or happens to be close to hadrons and thus fails the
isolation requirement. We further separate them
into clean and jetty events and then classify them by the number of
isolated leptons ($e$ and $\mu$). In addition to the acceptance cuts
described in Sec. 2, we impose an additional global requirement $\eslt >
40$~GeV, which together with the presence of jets, leptons or photons
may also serve as a trigger for these events.

Before proceeding to present results of our computation, we pause to
consider SM backgrounds to these events. We expect that the backgrounds
are smallest in the two photon channel, which we will mainly focus on
for the purpose of assessing the reach. We have not attempted to assess
the background because the recent analysis by the D0 collaboration
\cite{dzero}, searching for charginos and neutralinos in the GMSB
framework, points out that the major portion of the background arises
from mismeasurement of QCD jets and for yet higher values of $\eslt$
from misidentification of jets/leptons as photons. In other words, this
background is largely instrumental, and
hence rather detector-dependent. From Fig.~1 of Ref.~\cite{dzero}, we
estimate the inclusive $2\gamma+\eslt > 40$~GeV (60~GeV) background
level (for $E_T(\gamma_1,\gamma_2)>$ (20~GeV, 12~GeV)) to correspond to
$\sim$~0.9 (0.1) event in their data sample of $\sim 100$~$pb^{-1}$. The
background from jet mismeasurement, of course, falls steeply with
$\eslt$.  The inclusive $2\gamma+\eslt$ background is also
sensitive to the minimum $E_T$ of the photon.

To assess how changing the photon and $\eslt$ requirements alter the
SUSY signal, in Fig.~\ref{mlA:spec} we show the signal distribution of
({\it a})~$E_T(\gamma_2)$, the transverse energy of the softer photon in
two photon events, and ({\it b})~$\eslt$ in $\gamma\gamma+\eslt$ events
that pass our cuts, for three values of $\Lambda$. The following is
worth noting.

\begin{itemize}
\item For $\Lambda \simeq 100$~TeV (which we will see is in the range of
the Tevatron bound), reducing the $E_T(\gamma)$ cut does not increase
the signal. In fact, it may be possible to further harden this cut to
reduce the residual backgrounds. Although we have not shown it here, we
have checked that increasing the cut on the {\it hard} photon to
$E_T(\gamma_1) > 40$~GeV results in very little loss of signal for
$\Lambda > 100$~TeV.

\item In view of our discussion about SM backgrounds, it is clear that
requiring $\eslt > 60$~GeV greatly reduces the background with modest
loss of signal. Indeed, it may be possible to reduce the background to
negligible levels by optimizing the cuts on $E_T$ of the photons and on
$\eslt$.
\end{itemize}

The results of our computation of various topological cross sections at
a 2 TeV $p\bar{p}$ collider after cuts are shown in
Fig.~\ref{mlA:csection} for ({\it a}) 0 photon, ({\it b})~one photon,
and ({\it c})~two photon events. In this figure, we have required that
$\eslt > 60$~GeV. As mentioned, this reduces the cross section by just a
small amount, especially for the larger values of $\Lambda$ in this
figure. The solid lines correspond to cross sections for events with at
least one jet, while the dashed lines correspond to those for events
free of jet activity. The numbers on the lines denote the lepton
multiplicity, and are placed at those $\Lambda$ values that we
explicitly scanned. Finally, the heavy solid line represents the sum of
all the topologies, {\it i.e.} the inclusive SUSY cross section after
the cuts. We note the following.
\begin{itemize}
\item We have comparable signal cross sections in $1\gamma$ and $2\gamma$
channels. Since the background in the latter is considerably smaller
(recall a significant portion of it is from fake photons), the maximum
reach is obtained in the $2\gamma$ channel.

\item As anticipated, events with at least one jet dominate clean
events, irrespective of the number of photons.

\end{itemize}

We may obtain a conservative estimate of the reach by assuming an
inclusive $2\gamma + \eslt \geq 60$~GeV background level of 0.1 event
per 100~$pb^{-1}$; {\it i.e.} assuming a background level of
1~$fb$. This corresponds to a ``$5\sigma$ reach'' of 3.5~$fb$ (1~$fb$)
for an integrated luminosity of 2~$fb^{-1}$ (25~$fb^{-1}$) at the
Tevatron, or $\Lambda \leq 110$~TeV (130~TeV) at the Main Injector
(TeV33 upgrade).  As we have mentioned, it may be possible to further
reduce the background by hardening the $E_T(\gamma)$ and $\eslt$
requirements with only modest loss of signal. The background may also be
reduced if jet/lepton misidentification as a photon is considerably
smaller than in Run I~\cite{dzero}. If we optimistically assume that the
reach is given by the 5 (10) event level at the Main Injector (TeV33),
we would be led to conclude that experiments may probe $\Lambda$ values
as high as 118~TeV (145~TeV) at these facilities. It should be
remembered that $\Lambda=118$~TeV corresponds to $m_{\tg}\sim 950$~GeV,
almost equal to what is generally accepted as the qualitative upper limit
from fine tuning arguments.

We see from Fig.~\ref{mlA:prod}{\it b} that the $\tell_1\tell_1$
production cross section exceeds 1~$fb$ for $\Lambda \alt
100$~TeV. Since slepton production can lead to spectacular
$\ell\ell\gamma\gamma + \eslt$ events of the type observed by the CDF
Collaboration \cite{cdf}, it appears reasonable to ask whether signal
from slepton pair production might be observable at TeV33, and further,
whether it can be separated from a similar signal from chargino pair
production when each $\tw_1 \to \ell\nu\tz_1 \to
\ell\nu\tG\gamma$~\cite{footnote5}. The SM {\it physics} backgrounds
come from $WW\gamma\gamma$ production which for $E_{T\gamma} \geq
10$~GeV has a production cross section~\cite{cdf} of $0.15\pm
0.05$~$fb$, so that $\sim 0.1$ such event is expected in a data sample
of 25~$fb^{-1}$. The background from $t\bar{t}$ production is estimated
to be even smaller. In Fig.~\ref{mlA:slep} we show the total cross
section for clean $\ell\ell\gamma\gamma + \eslt$ events after cuts
(solid) and the corresponding cross section from just $\te_1$ and
$\tmu_1$ pair production (dashed). We see that for $\Lambda \alt
115$~TeV (corresponding to $m_{\tell_1} \sim 200
$~GeV), five or more
signal events should be present at TeV33, with about 60\% of these
having their origin in direct production of sleptons. Slepton pair
production alone yields five events for $m_{\tell_1} \alt 180$~GeV. If
instrumental backgrounds from jets faking an electron or photon turn out
to be negligible, direct detection of sleptons as heavy as 180~GeV may
be possible at TeV33~\cite{footnote6} for model line A. 
%Of course, a
%careful study is needed before definite conclusions can be made.
%THIS SEEMED REDUNDANT AND NOT VERY INFORMATIVE

\subsection{Model Line B: The Stau NLSP Scenario}

From Fig.~1, we see that we can obtain $\ttau_1$ as the NLSP for a wide
range of GMSB parameters. Here, we choose $n_5=2$, and take
$\tan\beta=15$ to make $\te_1$ and $\tmu_1$ somewhat heavier than
$\ttau_1$, with other parameters as before.  In Fig.~\ref{mlB:prod} we
show {\it a}) relevant sparticle masses, and {\it b}) cross sections for
the main sparticle production mechanisms versus $\Lambda$. For $\Lambda
\agt 30$~TeV, $m_{\ttau_1} \leq m_{\tz_1}$ but for $\Lambda \alt
32$~TeV, $\tz_1\to \tau\ttau_1$ is kinematically forbidden, and $\tz_1$
would decay via the four body decay $\tz_1 \to \nu_{\tau}\ttau_1 W^*$
(which is not yet included in ISAJET) or via its photon mode considered
above.  In our study, we only consider $\Lambda \geq 35$~TeV, the region
safe from LEP constraints.  Gluinos and squarks are then too heavy to be
produced at the Tevatron, and sparticle production is dominated by
chargino, neutralino and, to a lesser extent, slepton pair
production. 

The two body decay $\tw_1 \to \ttau_1 \nu$ is always
accessible, while the decay $\tw_1 \to W\tz_1$ becomes significant only
for $\Lambda \agt 45$~TeV ($m_{\tw_1} \agt 210$~GeV). The branching
fraction for $\tz_2$ decays are shown in Fig.~\ref{mlB:decay}{\it
a}. We see that $\tz_2$ decays via $\ttau_1 \tau$ with a branching
fraction that exceeds 0.5 if $m_{\tz_2} \alt 300$~GeV, and further that
branching fractions for $\tz_2\to \tell_1\ell$ ($\ell=e,\mu$) are not
negligible.  For the value of $\tan\beta$ in this Figure, the decay $\tz_2 \to
\tz_1 h$ is only important for relatively large values of $\Lambda$.
The lightest neutralino $\tz_1$ mainly decays via $\tz_1 \to
\ttau_1\tau$, though for large enough values of $\Lambda$ its decay to
sleptons of other families may also be significant.  The decay pattern
of the lighter selectron and smuon are illustrated in
Fig.~\ref{mlB:decay}{\it b}. For small values of $\Lambda$ (Region 2 of
Fig.~1), the decay $\tell_1 \to \ell \tz_1$ is kinematically allowed and
dominates. For larger values of $\Lambda$ (Region 4 of Fig.~1), where
this channel is closed, the neutralino is virtual and $\tell_1^- \to
\ttau_1^+ \ell^-\tau^-$ or $\tell_1^- \to \ttau_1^- \ell^-\tau^+$. These
decays dominate the decay $\tell_1 \to \ell\tG$.  The upshot of these
decay patterns is that SUSY events may contain several tau leptons from
sparticle cascade decays. At the very least, each event will contain a
pair of $\tau$s (in addition to other leptons, jets and $\eslt$) from
$\ttau_1$ produced at the end of the decay cascade. It is worthwhile to
note that the two $\tau$s could easily have the same sign of electric
charge.

The observability of SUSY realized as in this scenario thus depends on
the capability of experiments to identify hadronically decaying tau
leptons, and further, to distinguish these from QCD jets. Following the
same logic as in the $n_5=1$ case above, we now classify SUSY events by 
the number of identified taus, and further separate them into jetty and
clean event topologies labelled by the number of isolated leptons ($e$
and $\mu$). It should be remembered that the efficiency for identifying
taus is expected to be smaller than for identifying photons -- first, the
tau has to decay hadronically, and then the hadronic decay products have
to form a jet.

In our
analysis, in addition to the basic acceptance cuts discussed in Section
II,
we require $\eslt \geq 30$~GeV together with at least one of
the following which serve as a trigger for the events:
\begin{itemize}
\item one lepton with $p_T(\ell) \geq 20$~GeV,
\item two leptons each with $p_T(\ell) \geq 10$~GeV,
\item $\eslt \geq 35$~GeV.
\end{itemize}
In addition, we also impose the additional requirements:
\begin{itemize}
\item a veto on opposite sign, same flavor dilepton events with 
$M_Z-10~{\rm GeV} \leq  m(\ell\bar{\ell}) \leq M_Z + 10~{\rm GeV}$ to
remove backgrounds from $WZ$ and $ZZ$ and high $p_T$ $Z$ production, and 
\item for dilepton events, require $\Delta\phi(\ell\bar{\ell'}) \leq
150^{\circ}$ ($\ell,\ell'=e,\mu,\tau$) to remove backgrounds from $Z \to
\tau\bar{\tau}$ events.
\end{itemize}

The dominant physics sources of
SM backgrounds to $n$-jet + $m$-leptons + $\eslt$ events, possibly
containing additional taus, are $W$, $\gamma^*$ or $Z$ + jet
production, $t\bar{t}$ production and vector boson pair
production. Instrumental backgrounds that we have attempted to estimate
are $\eslt$ from mismeasurement of jet energy and mis-identification of
QCD jets as taus.

We have checked that even after these cuts and triggers, SM backgrounds
from $W$ production swamp the signal in channels with no leptons or just one
identified lepton ($e$, $\mu$ or $\tau$). The former is the canonical
$\eslt$ signal, which after optimizing cuts, may be observable at Run 2
if gluinos are lighter than $\sim 400$~GeV. We do not expect that this
signal from gluino and squark production will be detectable since even
for $\Lambda=35$~TeV, $m_{\tg} =578$~GeV with squarks somewhat
heavier. For this reason, and because there are large single lepton
backgrounds from $W$ production, we focus on signals with two or more
leptons in our study. Also, because the presence of $\tau$'s is the
hallmark of this scenario, we mostly concentrate on leptonic events with 
at least one identified $\tau$.

We begin by considering the signal and background cross sections for
clean events. These are shown in Table \ref{clean}. Events are
classified first by the number of identified taus, and then by the
lepton multiplicity; the $C$ in the topology column denotes
``clean'' events. For each topology, the first row of numbers denotes
the cross sections after the basic acceptance cuts and trigger
requirements along with the $Z$ veto and the $\Delta\phi$ cut discussed
above. We see that there is still a substantial background in several of the 
multilepton channels. This background can be strongly suppressed, with
modest loss of signal by
imposing an additional requirement,
\begin{itemize}
\item $p_{Tvis}(\tau_1) \geq 40$~GeV,
\end{itemize}
on the visible energy of the hardest tau in events with at least one
identified tau. In the background, the $\tau$s typically come from
vector boson decays, while in the signal a substantial fraction of these
come from the direct decays of charginos and neutralinos that are
substantially heavier than $M_Z$ (remember even
$m_{\tz_1}= 103$~(132)~GeV for $\Lambda =40$~(50)~TeV): thus signal taus
pass this cut more easily. A few points about this Table are worth
mentioning.
\begin{enumerate}
\item The signal cross sections in each channel are at most a few $fb$,
and with an integrated luminosity of 2~$fb^{-1}$, the individual signals
are below the $5\sigma$ level even for $\Lambda=40$~TeV. For the integrated
luminosity expected in Run II of the MI we will be forced to add the
signal in various channels and see if this inclusive signal is
observable.

\item The sum of the signal in all the channels in Table \ref{clean}, except
the $1\tau 1\ell$ channel which has a very large background, is shown in
the next two rows both with and without the $p_T$ cut on the $\tau$, while
in the last two rows we list $\sigma(sig)/\sqrt{\sigma(back)}$. 
We see that a somewhat better significance is obtained after the
$p_{Tvis}(\tau_1) \geq 40$~GeV cut.

\item We see that the inclusive SUSY signal in the clean channels for
the $\Lambda =40$~TeV case should be detectable with the Run II
integrated luminosity, whereas for the $\Lambda =50$~TeV case an
integrated luminosity of 12~$fb^{-1}$ is needed for a 5$\sigma$ signal.

\item We caution the reader that about 25-30\% of the $\tau$ background
comes from mis-tagging QCD jets as taus (except, of course, for the $W$
backgrounds and the backgrounds in the $C2\tau \ell$ channels
which are almost exclusively from these fake taus). Thus
our estimate of the background level is somewhat sensitive to the $\tau$
faking algorithm we have used. The signal, on the other hand, almost
always contains only real $\tau$s, so that improving the discrimination
between $\tau$ and QCD jets will lead to an increase in the projected
reach of these experiments.

\item In some channels the background is completely dominated by
fake taus. For instance, after the $p_{Tvis}(\tau)$ cut, the
$C1\tau 1\ell$ background from $W$ sources of just {\it
real taus} is only 1.9~$fb$, while the signal and other backgrounds
remain essentially unaltered from the cross sections in Table \ref{clean}.
Thus if fake $\tau$ backgrounds can be
greatly reduced, it may be possible to see a signal via other channels.

\end{enumerate}

Next, we turn to jetty signals for Model Line B. Cross sections for
selected signal topologies together with SM backgrounds {\it after} the 
$p_T(\tau_1) \geq 40$~GeV cut are shown in Table \ref{jetty}. The other
topologies appear to suffer from large SM backgrounds and we have not
included them here.

The following features are worth nothing.
\begin{enumerate}
\item We see that Model Line B results in smaller cross sections in
jetty channels. This should not be surprising since electroweak
production of charginos, neutralinos and sleptons are the dominant SUSY
processes, and because staus are light, branching fractions for hadronic
decays of $\tw_1$ and $\tz_{1,2}$ tend to be suppressed.

\item We see from Table \ref{jetty} that with the present set of cuts,
not only is the signal below the level of observability in any one of
the channels, but also that the inclusive signal is not expected to be
observable at the MI even for the $\Lambda = 40$~TeV case. With an
integrated luminosity of 25~$fb^{-1}$ the signal for the
$\Lambda=50$~TeV case is observable at the $6\sigma$ level.

\item As for the clean lepton case, a significant portion of the
background comes from QCD jets faking a tau. The fraction of events with
a fake tau varies from channel to channel, but for the $2\tau 1\ell$
channel in Table \ref{jetty} almost 60\% of the background (in contrast
to essentially none of the signal) involves at
least one fake $\tau$.

\item A major background to the jetty signal comes from $t\bar{t}$
production. To see if we could enhance the signal relative to this
background we tried to impose additional cuts to selectively reduce the
top background. Since top events are expected to contain hard jets, we
first tried to require $E_T(j) \leq 50$~GeV. We also, independently,
tried vetoing events where the invariant mass of all jets exceeded
70~GeV. While both attempts lead to an improvement of the signal to
background ratio, the statistical significance of the signal is not
improved (and is even degraded) by these additional cuts. We do not
present details about this for the sake of brevity. It may be possible to
reduce the top background by vetoing events with tagged $b$-jets,
but we have not attempted to do so here.

\end{enumerate}

For Model Line B, it appears that experiments at the MI should be able
to probe $\Lambda$ values up to just beyond 40~TeV in the inclusive clean
multilepton channels. It appears, however, that it will be essential to
sum up several channels to obtain a signal at the $5\sigma$
level. Confirmatory signals in inclusive jetty channels may be
observable at the $3.7\sigma$ level. Of course, for an integrated
luminosity of 25~$fb^{-1}$ it may be possible to probe $\Lambda=50$~TeV
even in the unfavoured jetty channels, and somewhat beyond in the clean
channels.  The situation is summarized in Fig.~\ref{mlB:reach}
where we show the signal cross sections summed over the selected channels
for events without jets (dashed) and for events with jets (solid). The
horizontal lines denote the minimum cross section needed for the signal
to be observable at the $5\sigma$ level, for the two assumptions about
the integrated luminosity.
We note that, in some channels, a susbtantial fraction of
background events come from QCD jets faking a tau --- our assessment of
the Run II reach is thus sensitive to our modelling of this jet mis-tag
rate. By the same token, if this rate can be reduced, the reach may be
somewhat increased. Finally, we remark that even though it appears that
the range of parameters that is accessible to experiments at the MI is
very limited ($\Lambda \leq 42$~TeV), these
parameters correspond to charginos as heavy as 192~GeV and gluinos and
squarks around 700~GeV.

\subsection{Model Line C: The co-NLSP Scenario}

This scenario can simply be obtained by choosing parameters in Regions 3
of Fig.~1, so that $\te_1$, $\tmu_1$
and $\ttau_1$ are all approximately degenerate, and $\te_1$ and $\tmu_1$
cannot decay to $\ttau_1$.  Here, we choose $n_5=3$ and $\tan\beta=3$
with other parameters as before.  In Fig.~\ref{mlC:prod} we show {\it
a}) relevant sparticle masses, and {\it b}) cross sections for the main
sparticle production mechanisms at the Tevatron versus $\Lambda$.  Aside
from the fact that lighter sleptons of all three flavours have
essentially the same mass, the main difference from the earlier model
lines that we have studied is that, while charginos and neutralinos
dominate for lower values of $\Lambda$, slepton pair production is the
dominant production mechanism for $\Lambda \agt 40$~TeV (corresponding
to $m_{\tell} \agt 125$~GeV).

In Fig.~\ref{mlC:decay} we show the branching fractions for {\it
a})~chargino, and {\it b}) neutralino decays versus $\Lambda$. For small
$\Lambda$ in frame {\it a}), charginos dominantly decay via $\tw_1 \to
\ttau_1\nu_{\tau}$ since the corresponding decays to smuons and
selectrons are suppressed by the lepton Yukawa coupling. As $\Lambda$
increases, decays to sneutrinos and the heavier (dominantly left-handed)
sleptons become accessible. Since these occur via (essentially
unsuppressed) gauge interactions, these rapidly dominate the decay to
$\ttau_1$. The decay $\tw_1 \to \tz_1 W$ also becomes significant for
$m_{\tw_1} \geq 200$~GeV. Turning to $\tz_2$ decays shown in frame {\it
b}), we see that these dominantly decay to sleptons with branching
fractions more or less independent of the lepton flavour. Again, since
$\tz_2$ is dominantly an $SU(2)$ gaugino, decays to the heavier
(dominantly left-handed) sleptons and sneutrinos dominate when these are
kinematically unsuppressed. The branching fraction for the decay $\tz_2
\to \tz_1 h$ is also significant. From the plot of sparticle masses in
Fig.~\ref{mlC:prod}{\it a}, we see that the heavier charged sleptons and
sneutrinos decay via $\tf_2 \to f\tz_1$ while $\tf_1 \to f\tG$ ($f
=\ell,\nu$). The lightest neutralino decays via $\tz_1 \to \tell_1 \ell$
with branching fractions essentially independent of the lepton flavour.

The bottom line of these decay patterns is that even though $\ttau_1$ is
strictly speaking the NLSP, we expect a large multiplicity of isolated
leptons ($e$ and $\mu$) from sparticle production at colliders. This is
because all flavours of sleptons are roughly equally produced in SUSY
cascade decays, and decays of $\te_1$ and $\tmu_1$ do not involve a stau
at an intermediate stage. This is illustrated in Fig.~\ref{mlC:mult} for
$\Lambda=30$~TeV and $\Lambda=40$~TeV, where we show the multiplicity
distributions for both $n_e+n_{\mu}$ and $n_e+n_{\mu}+n_{\tau}$, for
events satisfying the basic acceptance cuts (see Section~II)
and trigger conditions, but not the
additional requirements described in the Section~IIIB. We
see that while the lepton multiplicity is largest for two leptons (due
to production of $\ell_R$ pairs), a very sizeable fraction of signal
events have both $n_e+n_{\mu}$ and $n_e+n_{\mu}+n_{\tau} \geq 4$ for
which backgrounds from SM sources, shown in Table \ref{4lep}, are very
small. Here, for the $n_e+n_{\mu}+n_{\tau} \geq 4$ background sample, we
found that all the background events that passed the cuts automatically
satisfied $n_e+n_{\mu} \geq 2$. We also impose this requirement, which
should facilitate triggering on these events even without a $\tau$
trigger, on the signal~\cite{footnote7}.

In our simulation we found that the background from $W$ and $Z$ plus jet
production, as well as $t\bar{t}$ production are negligible. We see that
the $4$-lepton background is an order of magnitude smaller than the
corresponding $3\ell$ backgrounds in Tables \ref{clean} and \ref{jetty}
(even though the cuts there are more stringent than in Table
\ref{4lep}). In contrast, for the signal we see from Fig.~\ref{mlC:mult}
that the rate for 3$\ell$ events is much smaller than that for $\geq
4\ell$ events. Thus the $\geq 4\ell$ channel offers the best hope for
detection of the SUSY signal for model line C.

The SUSY reach for the co-NLSP model line is illustrated in
Fig.~\ref{mlC:reach} where we show the signal cross section versus
$\Lambda$ for inclusive events with $n_e+n_{\mu} \geq 4$ (dashed) and
$n_e+n_{\mu} +n_{\tau} \geq 4$ (solid) where we require, in addition,
$n_e+n_{\mu} \geq 2$. Here we have summed
the cross section for events with and without jets as this offers the
greatest reach. The corresponding horizontal lines denote the levels
where the signal will be just detectable at the `$5\sigma$ level' (with
a minumum of at least five events) at the MI and at the proposed TeV33
upgrade.  We observe the following.
\begin{itemize}
\item With an integrated luminosity of 2~$fb^{-1}$, the signal is
rate-limited in the $n_e+n_{\mu} \geq 4$ channel, and experiments at the
MI should be able to probe $\Lambda$ out to about 45~TeV (corresponding
to charginos heavier than 300 GeV) if we require a
minimum signal level of five events. Including taus increases the
signal but also increases the background so that the reach is only marginally
improved. Since this background largely comes from tau mis-identification,
it should be kept in mind that our projection for the reach via the
$n_e+n_{\mu} +n_{\tau} \geq 4$ channel is somewhat dependent on our
simulation of this.

\item For an integrated luminosity of 25~$fb^{-1}$ we see that the
projected increase in the background in the channel that includes taus
actually leads to a {\it reduction} of the reach, and the greatest reach
is obtained via events with $n_e+n_{\mu} \geq 4$ leptons for which the
background is very small. In our assessment of this reach we have
assumed that backgrounds from hadrons or jets mis-identified $e$s and
$\mu$s are negligible. The reach of TeV33 experiments should then extend
to $\Lambda \alt 55$~TeV which corresponds to $m_{\tw_1} (m_{\tg}) \alt
400 (1200)$~GeV!

\item Although we have not shown this here, we have checked that the
same sign dilepton channel does not yield a better reach than the
4$\ell$ channel just discussed. Typically, cross sections in this channel
are about 10-25\% of the total dilepton cross section in
Fig.~\ref{mlC:mult} whereas SM backgrounds (with just the basic cuts and
triggers) are in the several $fb$ range.
\end{itemize}

\subsection{Model Line D: The Higgsino NLSP scenario}

Within the GMSB framework described above, the value of $|\mu|$ that we
obtain tends to be considerably larger than $M_1$ and $M_2$, so that the
lightest neutralino is dominantly a gaugino (more specifically a bino,
since $M_1 \simeq \frac{1}{2}M_2$). This is indeed the case for the
three model lines examined above. Motivated by the fact that the
phenomenology is very sensitive to the nature of the NLSP, we have
examined a non-minimal scenario where we alter the ratio of $|\mu|/M_1$
by hand from its model value, and fix a small value of $|\mu|$ so that
the NLSP is mainly Higgsino-like. We do not attempt to construct a
theoretical framework which realizes such a scenario, but only mention
that the additional interactions needed to generate $\mu$ and also the
$B$-parameter in this framework could conceivably alter the
relation between $\mu$ and the gaugino masses. With this in mind, we use
ISAJET to simulate a light Higgsino scenario where we take $n_5=2$,
$\tan\beta=3$, $M/\Lambda =3$, $C_{grav}=1$ but fix $\mu
=-\frac{3}{4}M_1$ rather than the value obtained from radiative
electroweak symmetry breaking. In practice, we do so by using the weak scale
parameters obtained using the GMSB model in ISAJET as input parameters
for the MSSM model, except that we use $\mu =-\frac{3}{4}M_1$ at this
juncture. For this `small $\mu$' scenario we expect that the two
lightest neutralino and the lighter chargino will be Higgsino-like and
close in mass, while the heavier charginos and neutralinos will be
gaugino-like. The resulting spectrum is shown in
Fig.~\ref{mlD:prod}{\it a}. Indeed we see that $\tz_1$ is the NLSP
over the entire parameter range shown, and that $\tz_2$ and $\tw_1$ are
generally only 20-30 GeV heavier. As a result, the fermions from $\tw_1$
and $\tz_2$ decays to $\tz_1$ will be rather soft. Slepton masses are
essentially family-independent because $\tan\beta$ is small. The lighter
Higgs boson mass is just above 100~GeV, independent of $\Lambda$.

Sparticle production at the Tevatron is dominated by the production of
these Higgsino-like charginos and neutralinos as can be seen in
Fig.~\ref{mlD:prod}{\it b}. An important difference between this case
and chargino and neutralino production in the model lines examined above
is that $\tw_1\tz_1$ and $\tz_1\tz_2$ production is also
substantial. For a fixed chargino mass, however, the sparticle
production cross section is somewhat smaller for model line C than
it is for the other model lines.

We have already noted that fermions from the decays $\tw_1 \to
f\bar{f'}\tz_1$ and $\tz_2 \to f\bar{f}\tz_1$ are generally expected to be
soft so that signatures for
$\tz_i\tz_j$ or $\tz_i\tw_j$ production will closely resemble those for
$\tz_1\tz_1$ production. In other words, sparticle signatures in such a
scenario will be mainly determined by the $\tz_1$ decay pattern shown in
Fig.~\ref{mlD:decay}. For small values of $\Lambda$, $\tz_1 \to \tG
\gamma$. As $\Lambda$ increases, the decays $\tz_1 \to \tG Z$ and $\tz_1
\to \tG h$ become kinematically accessible, and the branching fraction
for the photon decay becomes unimportant, while the decay to the Higgs
boson becomes dominant. This is in sharp contrast to the gaugino NLSP
case where the decay to the Higgs scalar is strongly suppressed. 

For small values of $\Lambda$ (where $\tz_1$ mainly decays to via $\tz_1
\to \tG\gamma$), the strategy for extracting the SUSY
signal is as for model line A; {\it i.e.} to look for inclusive 2$\gamma
+ \eslt$ events. If we adopt the conservative background level of 1~$fb$
as in this study, a `$5\sigma$' reach is obtained at the MI (TeV33)
provided the signal cross section exceeds 3.5~$fb$ (1~$fb$). These
levels are shown as the horizontal dashed lines in
Fig.~\ref{mlD:reach} while the corresponding signal is shown by the
curve labelled $\sigma(\gamma\gamma)$. We see that at MI (TeV33)
experiments should be able to probe $\Lambda$ values out
to about 80~TeV (90~TeV) corresponding to $m_{\tw_1} \sim 165$~GeV
(180~GeV) via a search for $\gamma\gamma +\eslt$ events.

For larger values of $\Lambda$, the NLSP dominantly decays via $\tz_1
\to \tG h$ and the di-photon signal drops sharply. In this case, since
$h$ mainly decays via $h \to b\bar{b}$, the SUSY signal, which can
contain up to four $b$-quarks, will be characterized by multiple tagged
$b$-jet plus $\eslt$ events, which may also contain other jets, leptons
and possibly photons (if one of the NLSPs decays via the photon mode).
The dominant SM background to multi-$b$ events presumably comes from
$t\bar{t}$ production and is shown in Table \ref{bback}, where we have
also shown the signal cross section for $\Lambda=100$~TeV. For events
with one or two tagged $b$-jets, the $t\bar{t}$ backgrounds come when
the $b$s from $t$ decay are tagged; {\it i.e.} the rate for events where
other jets are mis-tagged as $b$'s is just a few percent. This is also
true for signal events. On the other hand, in the $3b$ channel at least
one of the tagged $b$s in the $t\bar{t}$ background has to
come from a $c$ or light quark or gluon jet that is misidentified as a
$b$-jet, or from an additional $b$ produced by QCD radiation. This is
not, however, the case for signal events which contain up to four
$b$-jets. In each of the last two columns of Table \ref{bback} where we
show the top background and the SUSY signal, we present two
numbers: the first of these is the cross section when all the tagged
jets come from real $b$'s, while the second number in parenthesis is the
cross section including $c$ and light quark or gluon jets that are
mistagged as $b$. Indeed we see that the bulk of the $3b$ background is
reducible and comes from mistagging jets, whereas the signal is
essentially all from real $b$s.

It is clear from Table \ref{bback} that the best signal to background ratio
is obtained for events with $\geq 3b$-jets. Our detailed analysis shows
that although the signal cross section is rather small,
$3b$-channel with a lepton veto (since top events with large $\eslt$
typically contain leptons) offers the best hope for identifying the
signal above SM backgrounds. We see that the
signal is of similar magnitude as the background for $\Lambda =100$~TeV,
a point beyond the reach via the $\gamma\gamma$ channel. To further
enhance the signal relative to the background we impose the additional
cuts,
\begin{itemize} 
\item $\eslt \geq 60$~GeV, and
\item 60~GeV $\leq m_{bb} \leq 140$~GeV for at least one pair of tagged
$b$-jets in the event.
\end{itemize}
The first of these reduces the signal from 2.7~$fb$ to 2.1~$fb$ while
the background is cut by more than half to 3.4~$fb$. The mass cut was
motivated by the fact that in these models, at least one pair of tagged
$b$'s comes via $h \to bb$ decay, with $m_h \sim 100$~GeV, while the
$b$'s from top decay form a continuum. We found, however, that this cut
leads to only a marginal improvement in the statistical significance and
the signal to background ratio. We traced this to the fact that, because
of top event kinematics, one $b$-pair is likely to fall in the `Higgs
mass window'. Reducing this window to $100 \pm 20$~GeV leads to a
slightly improved S/B but leads to too much loss of signal to improve
the significance.

The signal cross section via the $3b$ channel after the basic cuts as
well as the additional $\eslt$ and $m_{bb}$ cuts introduced above is
shown by the solid curve labelled $h \to bb$ in Fig.~\ref{mlD:reach}.
For small values of $\Lambda$, the signal is small because of the
reduction in the branching fraction for $\tz_1 \to h\tG$ decay.  The
corresponding dashed lines shows the minimum cross section for a signal
to be observable at the $5\sigma$ level at the MI and at TeV33. We see
that, at the MI, there will be {\it no observable} signal in this
channel. In fact, even for the $\Lambda$ value corresponding to the
largest signal the statistical significance is barely $2\sigma$, so that
a non-observation of a signal will not even allow exclusion of this model
line at the 95\% CL.  With 25~$fb^{-1}$ of integrated luminosity, the
signal exceeds the $5\sigma$ level for 82~TeV $\leq \Lambda \leq
105$~TeV (corresponding to $m_{\tw_1} \sim m_{\tz_1} \sim m_{\tz_2} \sim
220$~GeV), and somewhat extends the reach obtained via the
$\gamma\gamma$ channel. Furthermore there appears to be no window
between the upper limit of the $\gamma\gamma$ channel and the lower
limit of the $3b$-channel.  A few points are worth noting.
\begin{enumerate}
\item Since the background dominantly comes from events where a $c$ or
light quark or gluon jet is mis-tagged as a $b$-jet, the reach via the
$3b$ channel is very sensitive to our assumptions about the $b$ mis-tag
rate. Indeed, if the mis-tag rate is twice as big as we have assumed,
there will be no reach in this channel even at TeV33. 

\item The $3b$ signal starts to become observable for $\Lambda \agt
80$~TeV where the branching fraction for the decay $\tz_1 \to h\tG$
becomes comparable to that for the decay $\tz_1 \to \gamma\tG$. The
value of $\Lambda$ for which the Higgs decay of the neutralino becomes
dominant depends on $m_h$, which in turn is sensitive to
$\tan\beta$.

\item Although we have not shown it here, signals involving $b$-jets
together with additional photons or $Z$ bosons identified via their
leptonic decays have very small cross sections and appear unlikely to be
detectable even for $\Lambda \sim 100$~TeV.

\end{enumerate}

Despite the fact that the top background alone is 50 to several hundred
times larger than the SUSY signal in all relevant one and two tagged $b$
plus multilepton channels in Table \ref{bback}, we have examined whether
it was possible to separate the signal from the background.  We focussed
on the $2b + 0\ell$ signal which has the best $S/B$ ratio, and required
in addition that $\eslt \geq 60$~GeV (which reduces the background by
almost 50\% with about a 20\% loss of signal) and further 60~GeV $\leq
m_{bb} \leq$ 140~GeV (which reduces the background by another factor of
half with a loss of 25\% of the signal)~\cite{footnote8}. We found,
however, that the signal is below the 5$\sigma$ level over essentially
the entire range of $\Lambda$ even at TeV33: only for $\Lambda =80 \pm
5$~TeV does the signal cross section exceed this $5\sigma$ level of
7.7~fb. Moreover the $S/B$ ratio never exceeds about 15\% which falls
below our detectability criterion $S/B \geq 20$\%.  We found that while
it is possible to improve the $S/B$ ratio via additional cuts, these
typically degrade the statistical significance of the signal. We thus
conclude that for model line D, there will be no observable signal in
the $2b$ channel even at TeV33.

Before closing this discussion, it seems worth noting that we should
interpret the reach in Fig.~\ref{mlD:reach} with some care, because
unlike in the study of model lines A, B and C, we do not really have a
well-motivated underlying theory (that gives a Higgsino NLSP). We
realized this by arbitrarily taking $\mu = -\frac{3}{4}M_1$. The NLSP
decay pattern, and hence the precise value of the reach, would depend on
this choice which should be regarded as illustrative.
In general, however, for the Higgsino NLSP model line, the
coupling of the NLSP to Higgs bosons is substantial so that the
branching fraction for the decay $\tz_1 \to h\tG$ becomes large when this
decay is not kinematically suppressed. For small values of $\Lambda$,
such that
the NLSP can only decay via $\tz_1 \to \gamma \tG$, SUSY signals
should be readily observable in the $\gamma\gamma + \eslt$ channel; once
the NLSP decay to $h$ begins to dominate, the cross section for diphoton
events becomes unobservably small, and the SUSY signal mainly manifests
itself as multiple $b$ events which have large backgrounds from
$t\bar{t}$ production. The most promising way to search for SUSY then
seems to be via $\eslt$ events with $\geq 3$ tagged $b$-jets but for a
search in this channel an integrated luminosity of 25~$fb^{-1}$ appears
essential. A signal that extends the reach beyond that in the
$\gamma\gamma$ channel is possible provided experiments are able to
reduce the background from mis-tagged charm (light quark or gluon) jets
to below 5\% (0.2\%).

\section{Summary and Concluding Remarks}

The GMSB framework provides a phenomenologically viable alternative to
the mSUGRA model. The novel feature of this framework is that SUSY
breaking may be a low energy phenomenon. In this case, the gravitino is
by far the lightest sparticle, and the NLSP decays within the detector
into a gravitino and SM sparticles. Sparticle signals, and hence the
reach of experimental facilities, are then sensitive to the identity of
the NLSP.

In this paper, we have examined signals for supersymmetric particle
production at the Tevatron, and evaluated the SUSY reach of experiments
at the MI or at the proposed TeV33 within the GMSB framework. In our
study, we consider four different model lines, each of which lead to
qualitatively different experimental signatures. We use the event
generator to simulate experimental conditions at the Tevatron, and for
each model line, we have identified additional cuts that serve to
enhance the SUSY signal over SM backgrounds. We assume the NLSP decay is
prompt. This is a conservative assumption in that we do not make use of
a displaced vertex to enhance the signal over SM background.

For the first of these model lines, labelled A, the NLSP is mainly a
hypercharge gaugino and dominantly decays via $\tz_1 \to \tG\gamma$, so
that SUSY will lead to extremely striking events with multiple jets with
hard leptons and large $\eslt$ and up to two hard, isolated photons,
with cross sections (after all cuts) shown in
Fig.~\ref{mlA:csection}. The physics background to the $\gamma\gamma$
event topologies is very small, and detector-dependent instrumental
backgrounds (such as from jets being mis-identified as photons or
leptons, or large mismeasurement of transverse energies)
dominate~\cite{dzero}.  Even with a conservative estimate of 1~$fb$
for the background cross section, experiments at the MI (TeV33) should
be able to probe values of the model parameter $\Lambda$ out to 110~TeV
(130~TeV). If we optimistically assume that this background can be
reduced to a negligible level by hardening the cuts on the photons and
$\eslt$, we find a reach as high as $\Lambda \sim 118$~TeV
(corresponding to a gluino of 950~GeV) at the MI and of 145~TeV at
TeV33. For this model line, the $\ell\ell\gamma\gamma + \eslt$ signal
from slepton pair production may be observable at TeV33 even if sleptons
are as heavy as 180~GeV.

In model line B, the lighter stau is the NLSP, and heavier sparticles
cascade decayed down to the stau, which then decays via $\ttau_1 \to
\tau\tG$. The presence of isolated tau leptons~\cite{nandi}, in addition
to jets and other leptons is the hallmark of such a scenario. We found,
however, that in some channels, the background from misidentification of
QCD jets as $\tau$ completely swamp the physics backgrounds, making it
very difficult to detect the signal (see {\it e.g.} the $C1\tau1\ell$
channel in Table \ref{clean}) in this channel. We conclude that unless
$\tau$ misidentification can be greatly reduced from what we have
assumed, SUSY will only be detectable via channels with at least three
leptons ($e,\mu$ and $\tau$) for which the cross sections are
individually small, and then, only by summing the signal in several
channels. Even here, backgrounds from mis-identified taus are
significant. Our assessment of the reach is shown in
Fig.~\ref{mlB:reach}. We see that at the MI the clean multilepton
channel offers the best reach, out to $\Lambda = 42$~GeV (corresponding
to $m_{\tw_1}= 192$~GeV and squarks and gluinos as heavy as 700~GeV),
while at TeV33, the reach may be extended out to $\Lambda$ values
somewhat beyond 50~TeV ($m_{\tg} \sim 800$~GeV).
 
In model line C, the lighter stau is again the NLSP but the lighter
selectron and smuon are essentially degenerate with it, so that $\te_1$
and $\tmu_1$ cannot decay into a tau; {\it i.e.} these decay into a
gravitino and a corresponding lepton. Since cascade decays of sparticles
are equally likely to terminate in each flavour of slepton, we expect
that this model line will lead to very large multiplicities of $e$,
$\mu$ and $\tau$ in SUSY events. Indeed we found that the optimal
strategy in this case was to search for events with $n_e + n_{\mu} \geq
4$ or $n_e + n_{\mu} + n_{\tau} \geq 4$ with $n_e + n_{\mu} \geq 2$. The
reach is shown in Fig.~\ref{mlC:reach}. We see that at the MI, there
should be observable signals out to $\Lambda =45$~TeV while at TeV33
$\Lambda$ values as high as 55~TeV should be observable. These
correspond to a gluino (chargino) mass of 1000 (320)~GeV and 1200
(400)~GeV, respectively!

Finally, we have examined an unorthodox model line where, by hand, we
adjust the value of $\mu$ to be smaller than the value of the
hypercharge gaugino mass $M_1$. This leads to an NLSP which is
dominantly a Higgsino. Furthermore, $m_{\tw_1} \sim m_{\tz_2} \sim
m_{\tz_1}$ so that the fermions from $\tw_1$ and $\tz_2$ decays to
$\tz_1$ are soft, and SUSY event topologies are largely determined by
the decay pattern of $\tz_1$. For small values of $\Lambda$ for which
the NLSP can only decay via $\tz_1 \to \gamma \tG$, the signal is
readily observable in the $\gamma\gamma + \eslt$ channel. However, once
the NLSP decay to $h$ begins to dominate, SUSY mainly manifests itself
as multiple $b$ events which have large backgrounds from $t\bar{t}$
production. The most promising way to search for SUSY then seems to be
via $\eslt$ events with $\geq 3$ tagged $b$-jets and zero leptons. For a
search in this channel, an integrated luminosity of $\sim$25~$fb^{-1}$
appears essential. A signal that extends the reach beyond that in the
$\gamma\gamma$ channel is possible provided experiments are able to
reduce the background from mis-tagged charm (light quark or gluon) jets
to below 5\% (0.2\%). The reach for model line D is shown in
Fig.~\ref{mlD:reach}, but it should be kept in mind that the details of
this figure will be sensitive to our assumption about $\mu$.

To conclude, in GMSB models signals for SUSY events will be
quantitatively and qualitatively different from those in the mSUGRA
framework. This is primarily because a neutralino NLSP decays into a
photon, a $Z$ boson or a Higgs boson and a gravitino, or sparticles
cascade decay to a slepton NLSP which decays to leptons and a gravitino:
these additional bosons (or their visible decay products), or leptons
from slepton NLSP decays, often provide an additional handle which may be
used to enhance the signal over SM background. Although we have not
performed an exhaustive parameter scan, for the model lines that we
studied we found that the SUSY reach (as measured in terms of the mass
of the dominantly produced sparticles) is at least as big, and
frequently larger than in the mSUGRA framework. For some cases, this
conclusion depend on the capability of experiments to identify $\tau$
leptons and $b$-quarks with moderately high efficiency and purity. In
view of the diversity of signals that appear possible for just this one
class of models, we encourage our experimental colleagues to be in
readiness for tagging third generation particles as they embark on the
search for new phenomena in Run II of the Tevatron.

%%%%%%%%%%%%%%%%%%%%%%%%% ACKNOWLEDGEMENTS %%%%%%%%%%%%%%%%%%%%%%%%%%%%%%%%%%%%%
%
%\newpage
\acknowledgments We are grateful to our colleagues in the Gauge-mediated
SUSY breaking Group of the Run II SUSY and Higgs Workshop, especially
Steve Martin, Scott Thomas, Ray Culbertson and Jianming Qian for sharing
their insights. Model lines I and II were first studied at this
Workshop. We thank Regina Demina for her guidance on $b$-jet tagging and
mistagging efficiencies.  P.M. was partially supported by 
Funda\c{c}\~ao de Amparo \`a Pesquisa do Estado de
S\~ao Paulo (FAPESP). This research was supported in part by the
U.~S. Department of Energy under contract number DE-FG02-97ER41022 and
DE-FG-03-94ER40833. 
%
%%%%%%%%%%%%%%%%%%%%%%%%%%%% APPENDIX %%%%%%%%%%%%%%%%%%%%%%%%%%%%%%%%%%%%%%%%%%

%
%\appendix{\ \ SINGLE PHOTON BREMSSTRAHLUNG}
%\newpage
%
%%%%%%%%%%%%%%%%%%%%% REFERENCES %%%%%%%%%%%%%%%%%%%%%%%%%%%%%%%%%%%%%%%%%%%%%%
%

%%%%%%%%%%%%%%%%%%%%%%%%%%%% TABLES %%%%%%%%%%%%%%%%%%%%%%%%%%%%%%%%%%%%%%%

\begin{table}
\caption[]{SM background cross sections in $fb$ for various clean
multilepton topologies from $W$, $Z \to \tau\tau$, $VV$ ($V=W,Z$) and
$t\bar{t}$ production at a 2~TeV $p\bar{p}$ collider, together with
signal cross sections for $\Lambda=40$~TeV and $\Lambda= 50$~TeV for
Model Line B described in the text. For each event topology, the first
number denotes the cross section after the basic acceptance cuts and
trigger requirements along with the $Z$ veto and the $\Delta\phi$ cut
discussed in the text. The second number is after the additional cut,
$p_{Tvis}(\tau_1) \geq 40$~GeV, for events at least one identified
$\tau$. The entries labelled $Total^*$ are the sum of all the cross
sections except those in the $1\tau 1\ell$ channel.  The last two rows
provide a measure of the statistical significance of the signal.}
\label{clean}
\bigskip
\begin{tabular}{|c|cccc|cc|}
Topology & $W$ & $Z\to \tau\tau$ & $VV$ & $t\bar{t}$ & $\Lambda=40$~TeV
&$\Lambda=50$~TeV \\
\tableline
$C3\ell$ & 0 & 0 & 0.39 & 0 & 0.68 & 0.24 \\
         & 0 & 0 & 0.39 & 0 & 0.68 & 0.24 \\
$C1\tau 1\ell$ & 1045 & 4.2 & 36 & 0.044 & 8.6 & 1.96 \\
               &  43  & 2.0 & 10.8 & 0 & 5.3 & 1.27 \\
$C1\tau 2\ell$ & 0 & 0.57 & 1.4 & 0 & 3.3 & 0.93 \\
               & 0 & 0.045 &    0.43 & 0  &  1.9  & 0.59 \\
$C1\tau 3\ell$ & 0 & 0 & 0 & 0 & 0.31 & 0.16  \\
               & 0 & 0 & 0 & 0 & 0.21 & 0.10  \\
$C2\tau 1\ell$ & 0 & 1.5 & 1.2 & 0 & 4.1 & 1.2 \\
               & 0    & 0.57 & 0.79 &0 & 3.3 & 1.02 \\
$C2\tau 2\ell$ & 0 & 0 & 0 & 0 & 0.36 & 0.23 \\
               &  0 & 0 & 0 & 0 & 0.33 & 0.22 \\
\hline
$Total^*$      & 0 & 2.1 & 2.99 &  0 & 8.75 & 2.76 \\ 
               & 0   & 0.62 & 1.61 & 0 & 6.42 & 2.17 \\
\hline
$\sigma(sig)/\sqrt{\sigma(back)}$~($fb^{1/2}$) & & & & &3.87 & 1.22\\
                        & & & & & 4.30 & 1.45
\end{tabular}
\end{table}

%%%%%%%%
%%%%%%%%
 
\begin{table}
\caption[]{SM background cross sections in $fb$ for various jetty
multilepton topologies from $W$, $Z\to \tau\tau$, $VV$ ($V=W,Z$) and
$t\bar{t}$ production at a 2~TeV $p\bar{p}$ collider, together with
signal cross sections for $\Lambda=40$~TeV and $\Lambda= 50$~TeV for
Model Line B described in the text. The cross sections are with all the
cuts including the $p_{Tvis}$ cut on the hardest $\tau$.}

\label{jetty}
\bigskip

\begin{tabular}{|c|cccc|cc|}
Topology & $W$ & $Z\to \tau\tau$ & $VV$ & $t\bar{t}$ & $\Lambda=40$~TeV
&$\Lambda=50$~TeV \\
\tableline
$J3\ell$ & 0 & 0.019 & 0.28 & 0.3 & 1.06 & 0.35 \\
$J1\tau 2\ell$ & 0 & 0.19 & 0.29 & 1.2 & 1.92 & 0.79 \\
$J2\tau 1\ell$ & 0.11 & 0.79 & 0.41 & 0.8 & 2.25 & 1.18 \\
$J2\tau 2\ell$ & 0 & 0 & 0 & 0 & 0.31 & 0.22 \\
\hline
$Total$    & 0.11  & 1.0 & 0.98 & 2.3 & 5.53& 2.54 \\
\hline
$\sigma(sig)/\sqrt{\sigma(back)}$~($fb^{1/2}$) & & & & &2.64 & 1.21

\end{tabular}
\end{table}

%%%%%%%%
%%%%%%%%

\begin{table}
\caption[]{SM background cross sections in $fb$ for clean and jetty
events with $\geq 4$~leptons from $VV$ ($V=W,Z$) production at a 2~TeV
$p\bar{p}$ collider, after the basic cuts and triggers described in the
text. At least 2 leptons are required to be $e$ or $\mu$.
Backgrounds from $Z$, $W$ and $t\bar{t}$ production are
negligible. Also shown are corresponding signal cross sections for
$\Lambda=30$~TeV and $\Lambda= 40$~TeV for Model Line C described in the
text. As before, the $C$ ($J$) refers to clean and jetty events.}

\label{4lep}
\bigskip

\begin{tabular}{|c|c|cc|}
Topology & $VV$ & $\Lambda=30$~TeV &$\Lambda=40$~TeV \\
\tableline
$C:n_e+n_{\mu}\geq 4$ & 0.09 & 14.0 &  2.3  \\
$C:n_e+n_{\mu}+ n_{\tau}\geq 4$ & 0.30 & 19.0 & 3.0  \\
$J:n_e+n_{\mu}\geq 4$& 0 & 16.5 & 3.4 \\
$J:n_e+n_{\mu}+ n_{\tau}\geq 4$ & 0.33 & 21.2 & 4.5 

\end{tabular}
\end{table}

%%%%%%
%%%%%%
\begin{table}
\caption[]{The background cross section in $fb$ for multiple tagged
$b$-jets plus lepton plus $\eslt$ events from $t\bar{t}$ production
after basic cuts and triggers discussed in the text. Also shown are the
corresponding SUSY signal cross sections for $\Lambda =100$~TeV for
model line D. The numbers in parenthesis for the $3b$-channel include
events from charm or light quark jets jets faking a $b$-jet. Whereas
this fake rate dominates the background in the $3b$ channel, it is
negligible in the $1b$ and $2b$ channels.}

\label{bback}
\bigskip

\begin{tabular}{|c|cc|cc|cc|}
 & \multicolumn{2}{c |} {$1b$ } &\multicolumn{2}{c |}{$2b$ } & 
\multicolumn{2}{c |} {$\geq 3b$ } \\ 
%$1b$ & $2b$ & $2b$ & $3b$ & $3b$ \\ 
%
 &  $t\bar{t}$ & $\Lambda =100$~TeV & $t\bar{t}$ & $\Lambda =100$~TeV & 
$t\bar{t}$ & $\Lambda =100$~TeV  \\
\tableline
$0\ell$ &  508 & 11.6 & 221 & 7.5 & 1.2 (8.1) & 2.6 (2.7)\\       
$1\ell$ & 812 & 1.2 & 345 & 0.57 & 1.5 (9.3) & 0.11 (0.11)  \\
$2\ell$ & 132 & 0.49 & 56 & 0.31 & 0.13 (0.31) & 0.04 (0.05) \\
$3\ell$ & 0.22 & 0.03 & 0 & 0.04 & 0 (0) & 0 (0)
%
%\hline

\end{tabular}
\end{table}

%%%%%%%%%%%%%%%%%%%%%% FIGURE CAPTIONS %%%%%%%%%%%%%%%%%%%%%%%%%%%%%%%%%%%%%%
\iftightenlines\else\newpage\fi
\iftightenlines\global\firstfigfalse\fi
\def\dofig#1#2{\iftightenlines\epsfxsize=#1\centerline{\epsfbox{#2}}\bigskip\fi}

%FIG. 1
\begin{figure}
\dofig{7in}{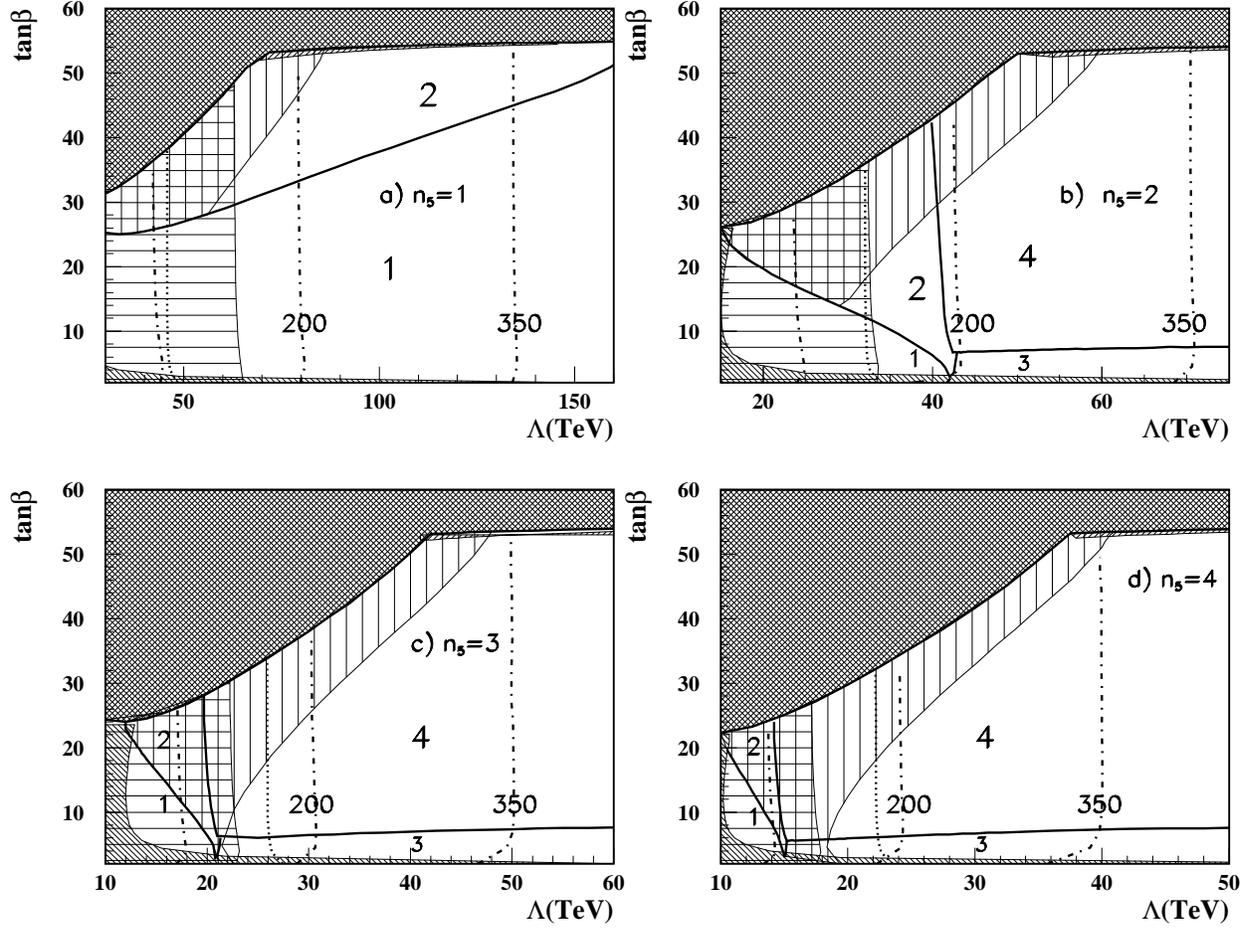}
\caption[]{The four regions of the $\Lambda-\tan\beta$ parameter plane
discussed in Sec.~I of the text. We fix $M=3\Lambda$, and take $\mu$ to
be positive. The heavy solid lines denote the
boundaries between these regions. The grey region is excluded because
electroweak symmetry is not correctly broken. The shaded regions should
be probed by experiments at LEP and are nearly excluded because
$m_{\ttau} \leq 76$~GeV (vertical shading), $\tz_1 \leq 84$~GeV
(horizontal shading) or $m_h \leq 95$~GeV (or $m_A \leq 83$~GeV)
(diagonal shading). The dot-dashed contours are where the chargino mass
is 95, 200 or 350~GeV, while the dotted line is the contour of
$m_{\te_1} = 90$~GeV.}
\end{figure}

%FIG. 2
\begin{figure}
\dofig{7in}{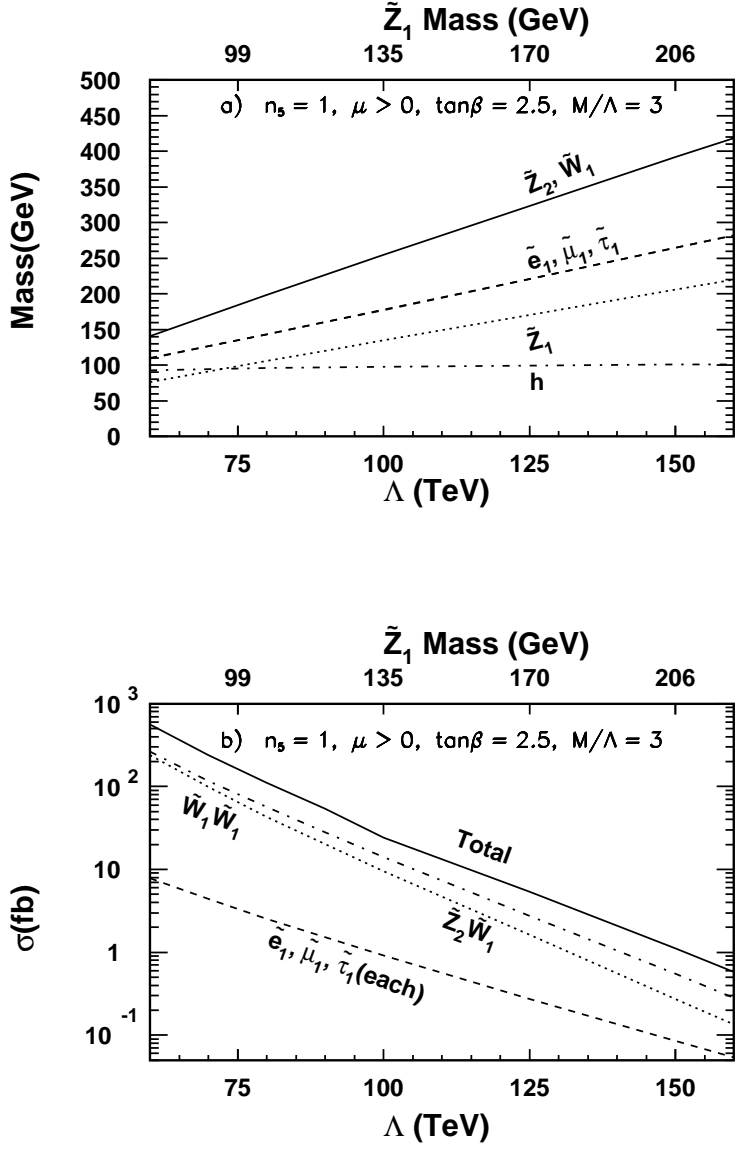}
\caption[]{{\it a}) Relevant sparticle masses, and {\it b}) production
cross sections for the main sparticle production reactions at a 2~TeV
$p\bar{p}$ collider versus the parameter $\Lambda$ for the bino
NLSP model line A. In frame  {\it b}) the dot-dashed line represents the 
chargino pair production, while the dotted line denotes that for
$\tw_1\tz_2$ production. Also shown on the upper axis is the mass of the
NLSP.}
\label{mlA:prod}

\end{figure}

%FIG. 3
\begin{figure}
\dofig{6in}{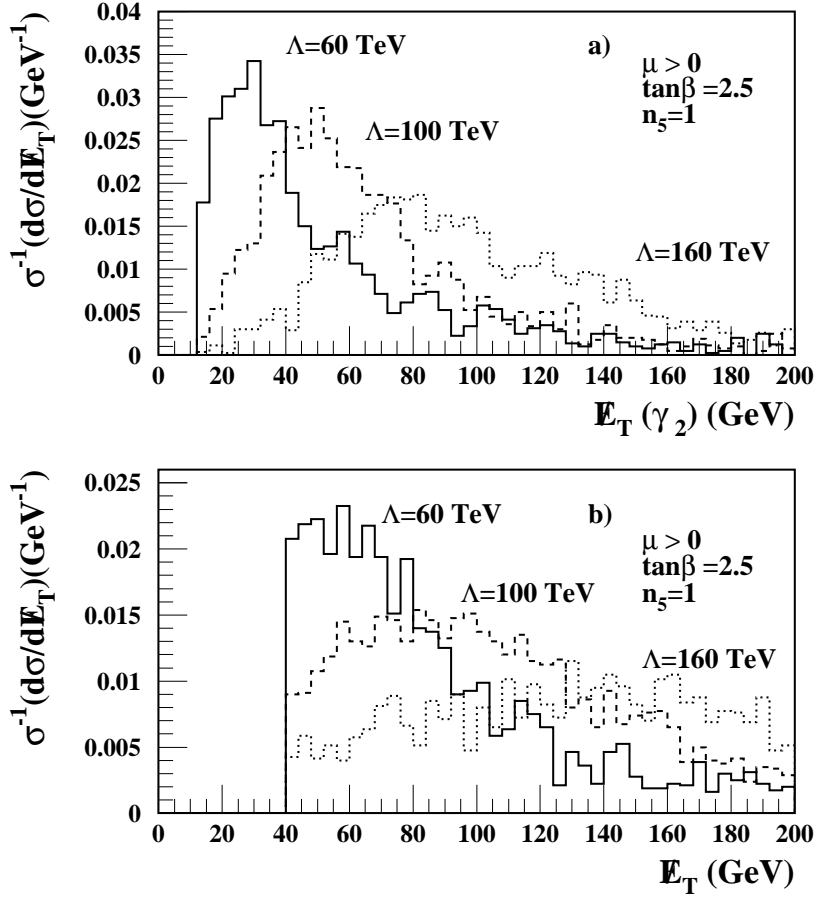}
\caption[]{{\it a})~The transverse energy distribution for the softer
photon , and {\it b}) the $\eslt$ distribution for the inclusive
$\gamma\gamma+\eslt$ for model line A.}
\label{mlA:spec}
\end{figure}

%FIG. 4 
\begin{figure}
\dofig{6in}{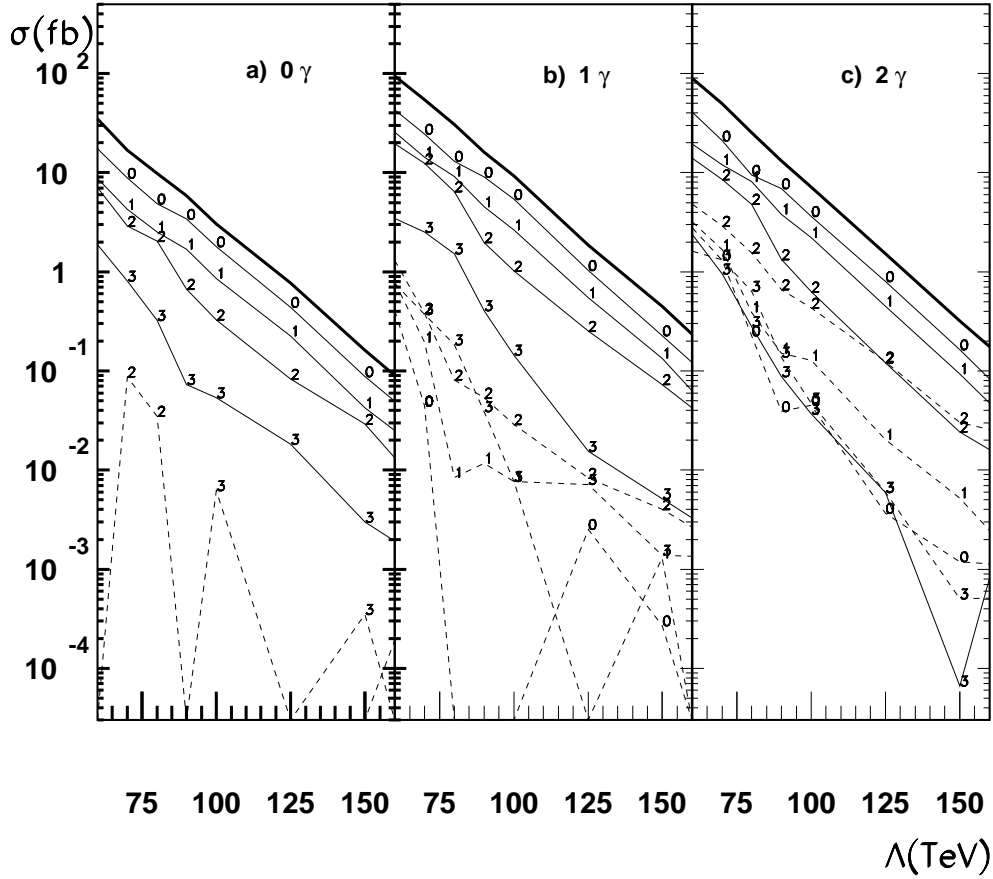}
\caption[]{Topological cross sections for inclusive {\it a}) 0$\gamma$, {\it
b})~$1\gamma$, and {\it c})~$\gamma\gamma$ plus $\eslt$ signals after
the cuts and triggers discussed in the text for model line A. The solid
lines denote cross sections for events with at least one jet, while the
dashed lines denote cross sections for events with no jets. The numbers
on the lines denote the lepton multiplicity. The heavy solid line
denotes the total signal cross section after all the cuts.}
\label{mlA:csection}
\end{figure}

%FIG. 5
\begin{figure}
\dofig{6in}{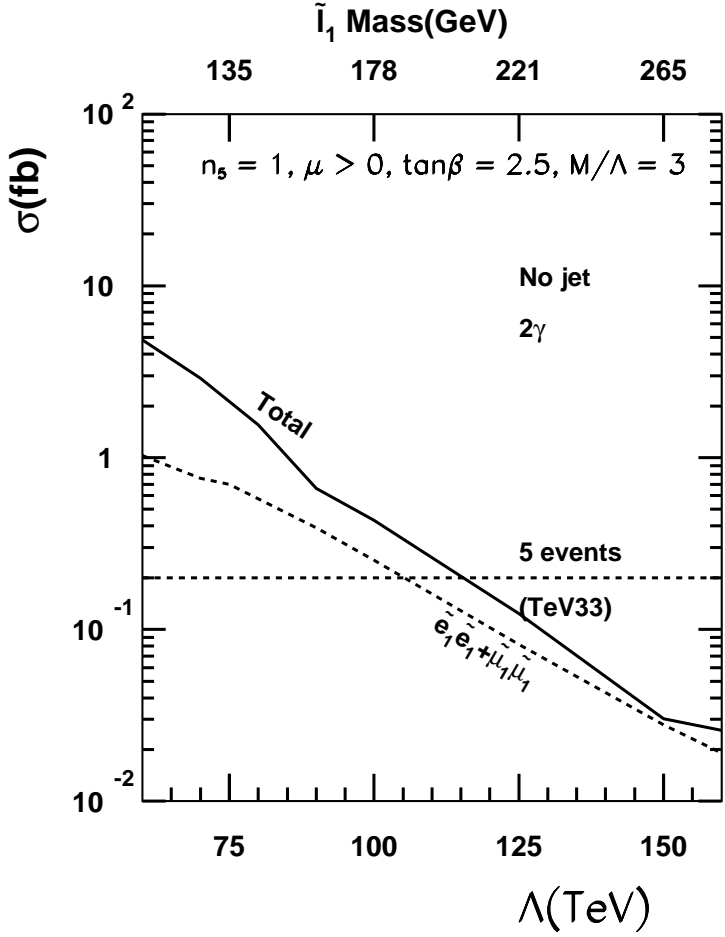}
\caption[]{Signal cross sections for clean $\ell\ell\gamma\gamma +\eslt$
events ($\ell =e,\mu$) from all SUSY sources (solid), and from just
$\te_1$ or $\tmu_1$ pair production (dashed) versus $\Lambda$ for model
line A. The upper scale denotes $m_{\tell_1}$.}
\label{mlA:slep}
\end{figure}

%FIG. 6
\begin{figure}
\dofig{5in}{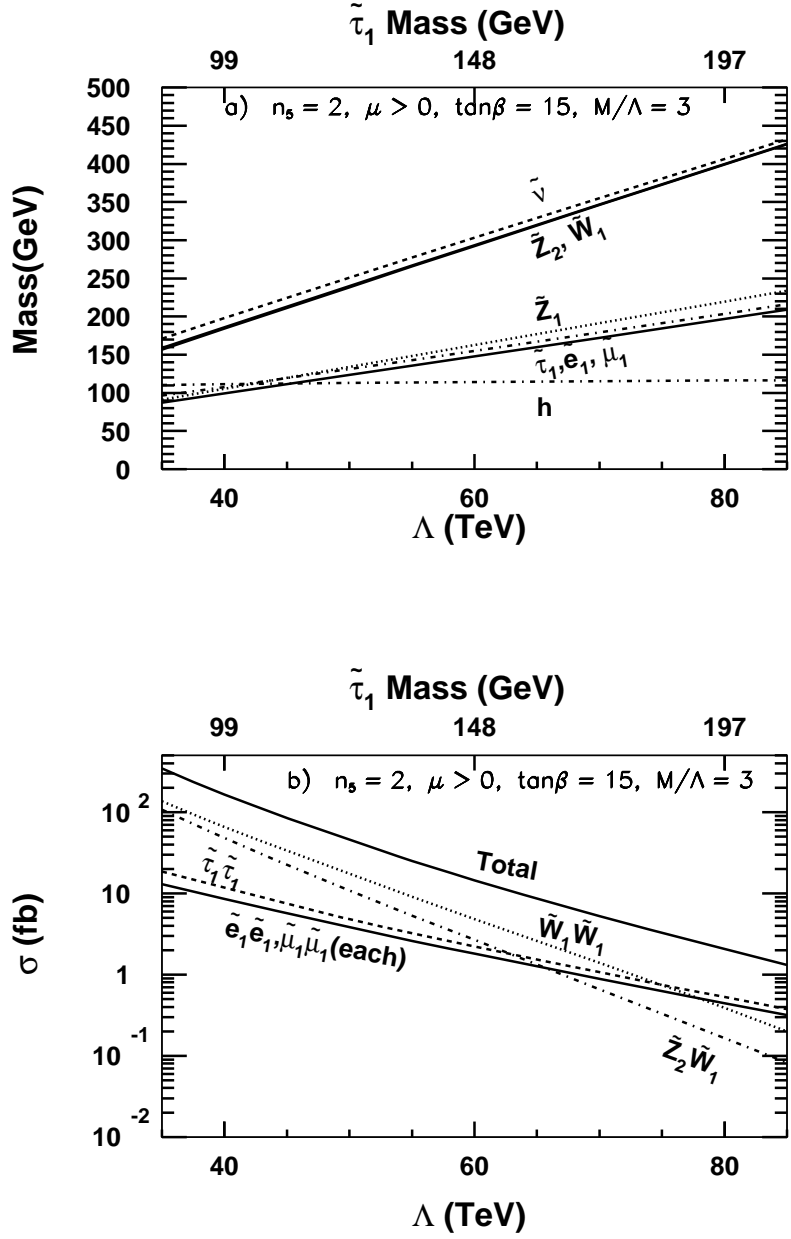}
\caption[]{The same as Fig.~\ref{mlA:prod}, but for the stau NLSP model
line B. In frame {\it a}) the solid line denotes the mass of $\ttau_1$
while the dot-dashed line denotes the mass of the lighter selectron or
smuon which are very nearly degenerate.}
\label{mlB:prod}
\end{figure}

%FIG. 7
\begin{figure}
\dofig{7in}{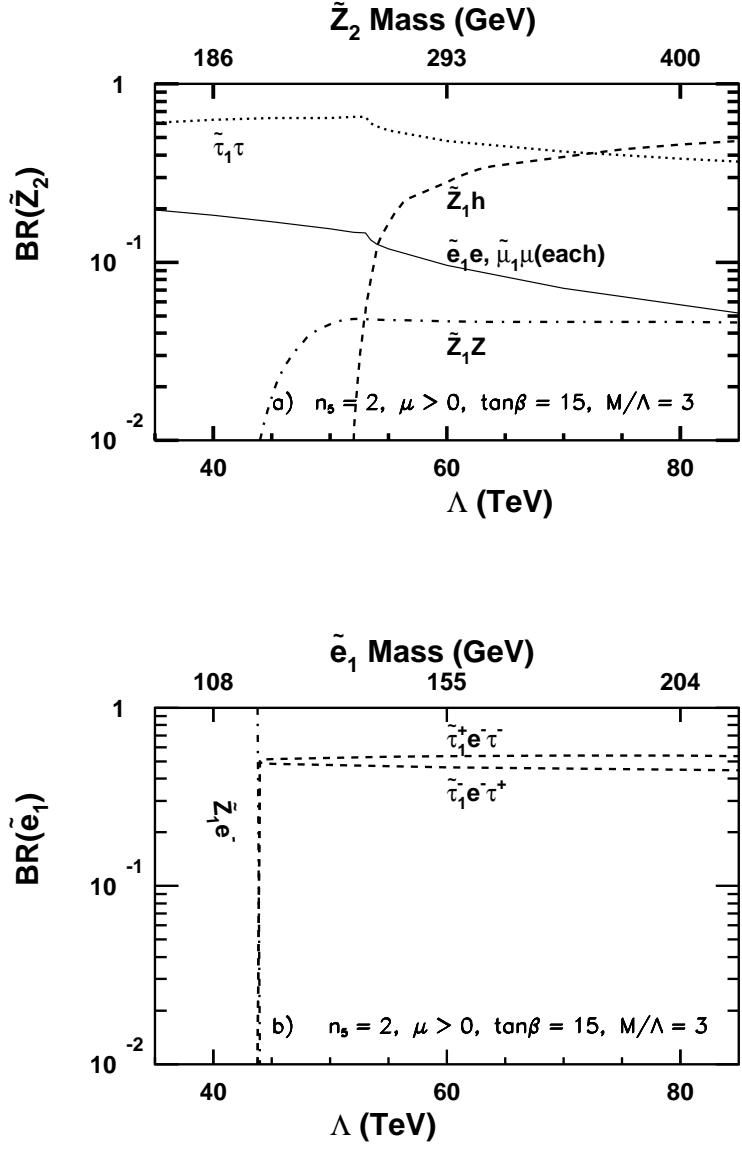}
\caption[]{The branching ratios for various decays of {\it
a})~$\tz_2$, and {\it b})~$\tell_1$ for model line B. The upper scale
shows the mass of the parent sparticle. The relevant decay
patterns of other sparticles are described in the text.}
\label{mlB:decay}
\end{figure}

%FIG. 8
\begin{figure}
\dofig{6in}{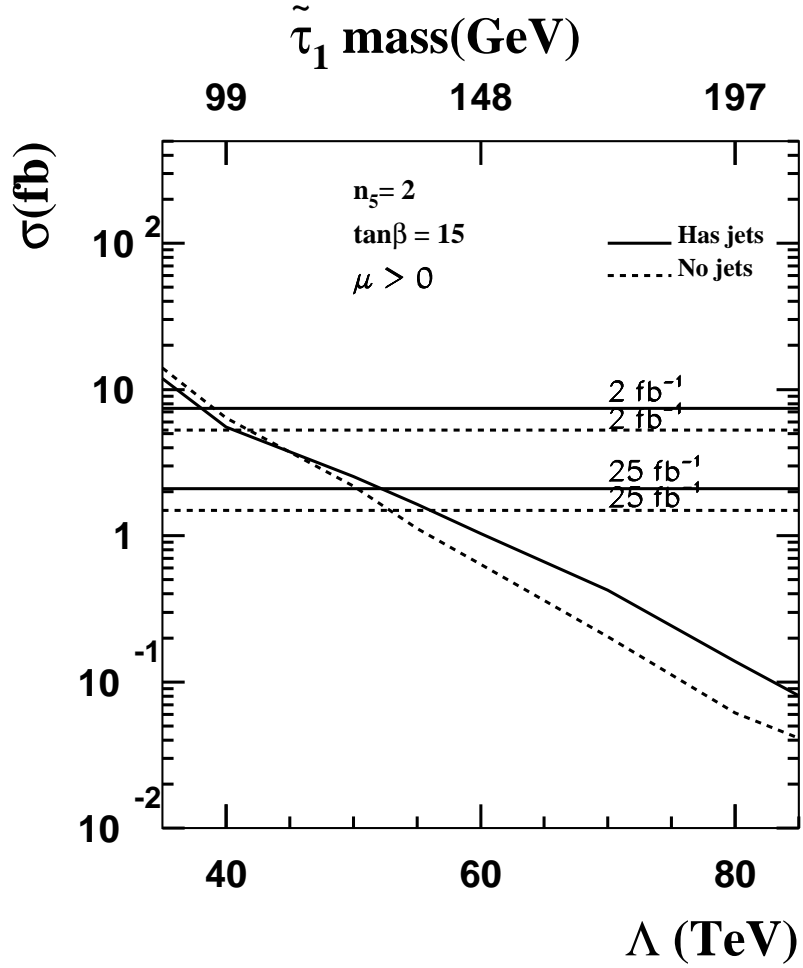}
\caption[]{Signal cross sections after all cuts versus $\Lambda$ for the
total signal in the clean (dashed line) and jetty (solid line) channels
shown in Table~\ref{clean} and Table~\ref{jetty}, respectively, for model
line B. The corresponding horizontal lines denote the minimum cross
section for a $5\sigma$ signal, both at the MI and at TeV33.}
\label{mlB:reach}
\end{figure}

%FIG. 9
\begin{figure}
\dofig{7in}{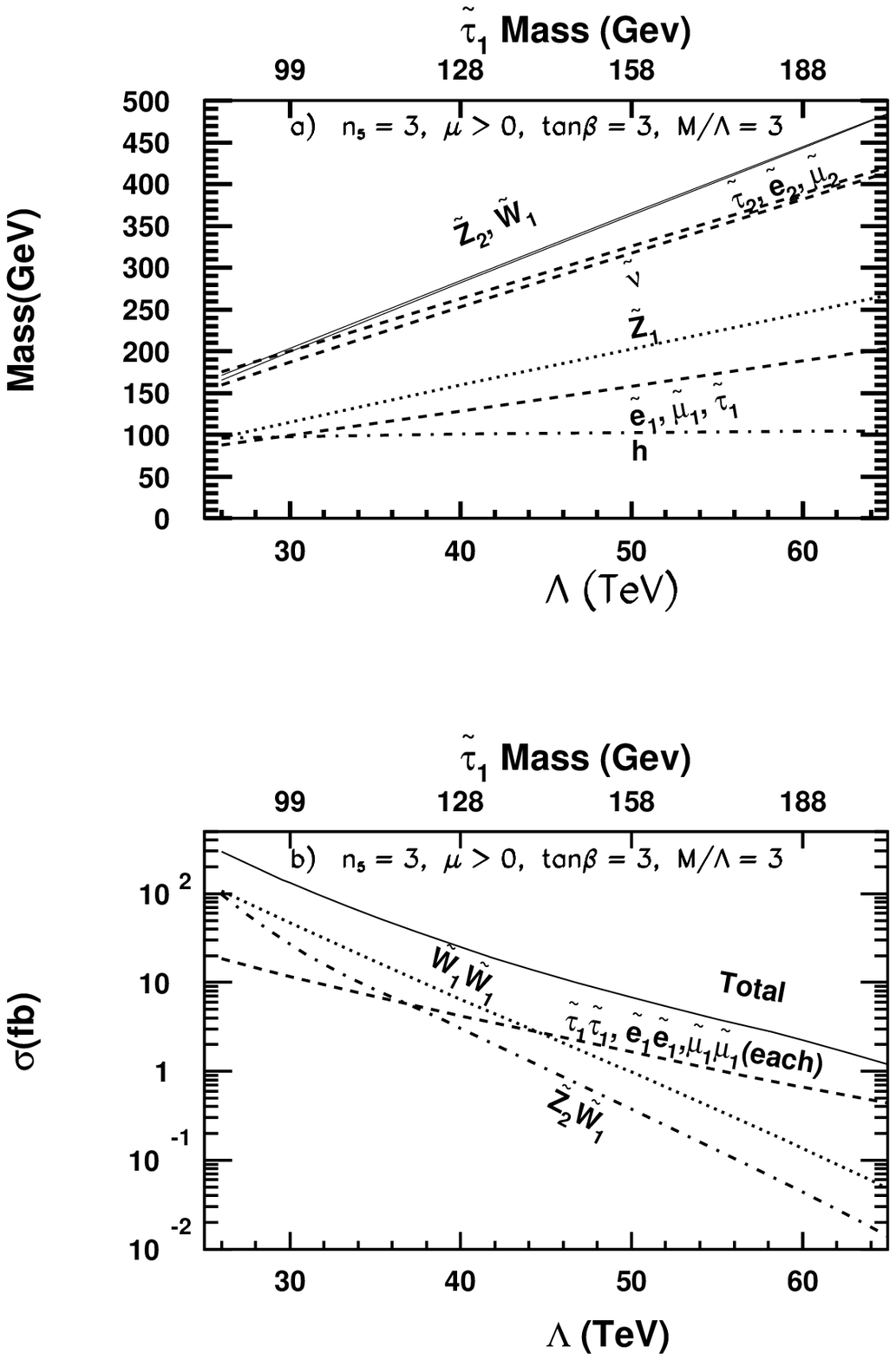}
\caption[]{The same as Fig.~\ref{mlA:prod}, but for the co-NLSP model line C.}
\label{mlC:prod}
\end{figure}

%Fig. 10
\begin{figure}
\dofig{7in}{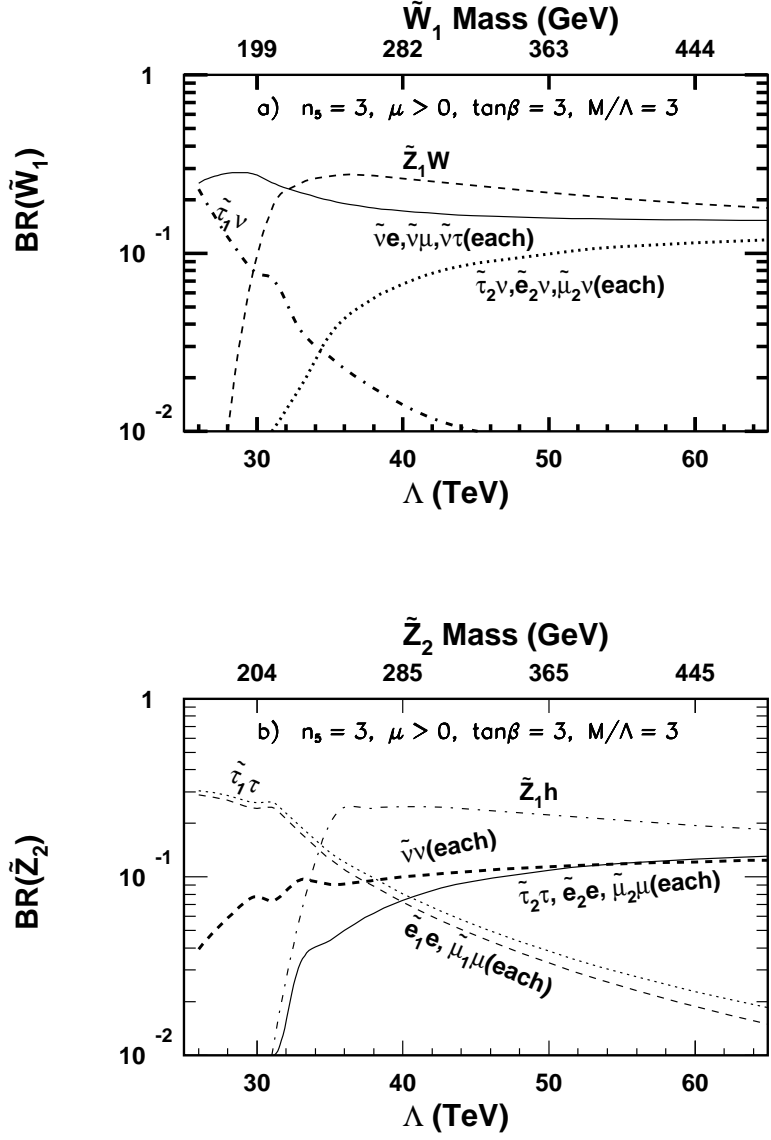}
\caption[]{Branching ratios for decays of {\it a})~$\tw_1$ and {\it
b})~$\tz_2$ for model line C. Decay patterns of other sparticles are
discussed in the text.}
\label{mlC:decay}
\end{figure}

%Fig. 11
\begin{figure}
\dofig{6in}{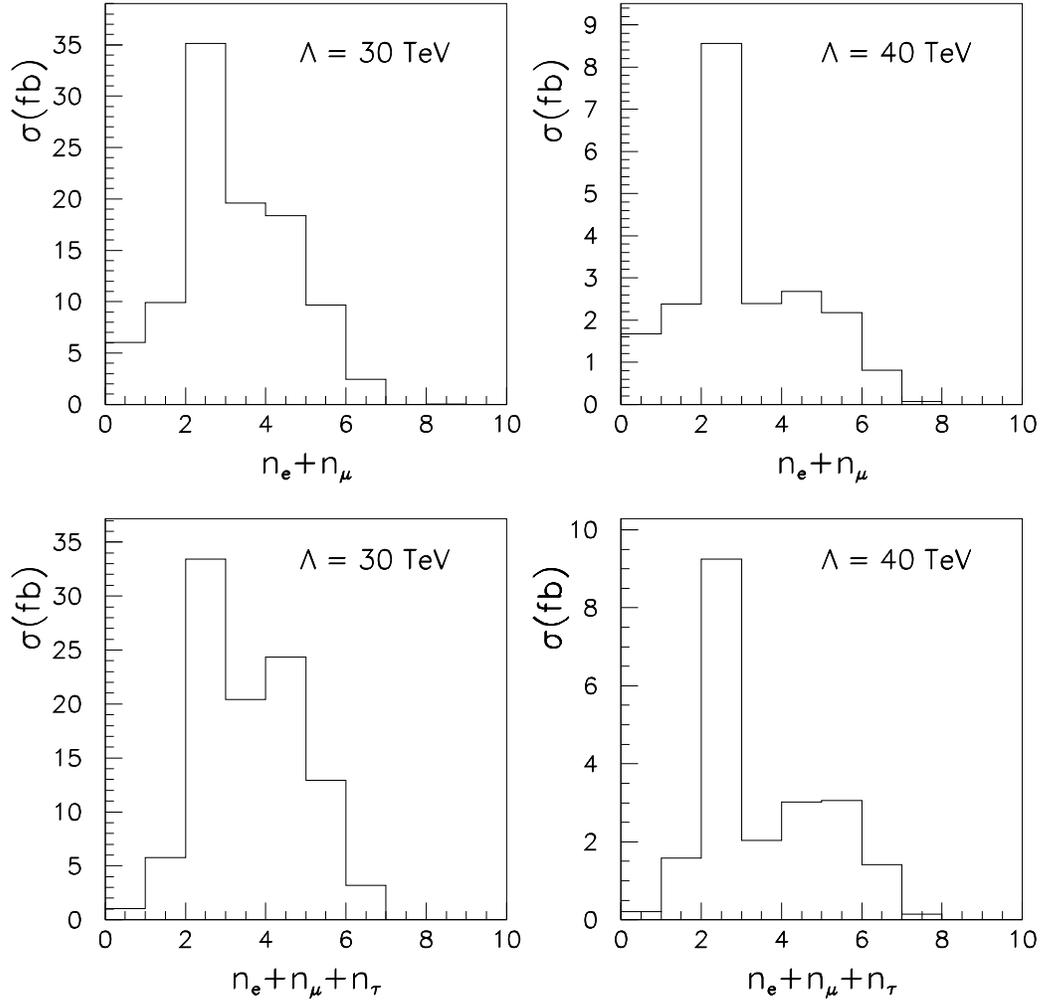}
\caption[]{Lepton multiplicity distributions for SUSY events after all
cuts with $\Lambda = 30$~TeV, and $\Lambda = 40$~TeV for model line
C. We show distributions of both $n_e+n_{\mu}$ and
$n_e+n_{\mu}+n_{\tau}$, where $n_{\tau}$ is the multiplicity of tagged
$\tau$ leptons in the event.}
\label{mlC:mult}
\end{figure}

%Fig. 12
\begin{figure}
\dofig{6in}{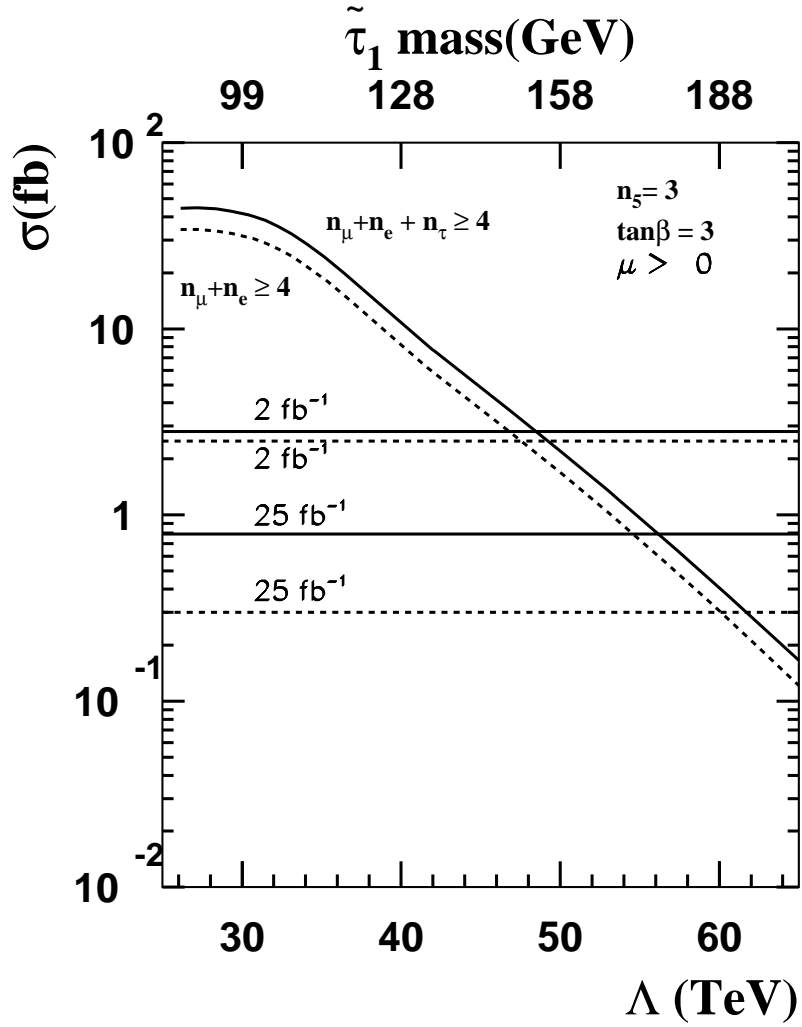}
\caption[]{Signal cross sections after all cuts for SUSY events with
$n_e+n_{\mu}+n_{\tau} \geq 4$ (solid) and $n_e + n_{\mu} \geq 4$
(dashed) for model line C. The corresponding horizontal lines denote the
minimum cross section for a $5\sigma$ signal (with a minimum of five
signal events) at the MI and at TeV33.}
\label{mlC:reach}
\end{figure}

%Fig. 13
\begin{figure}
\dofig{7in}{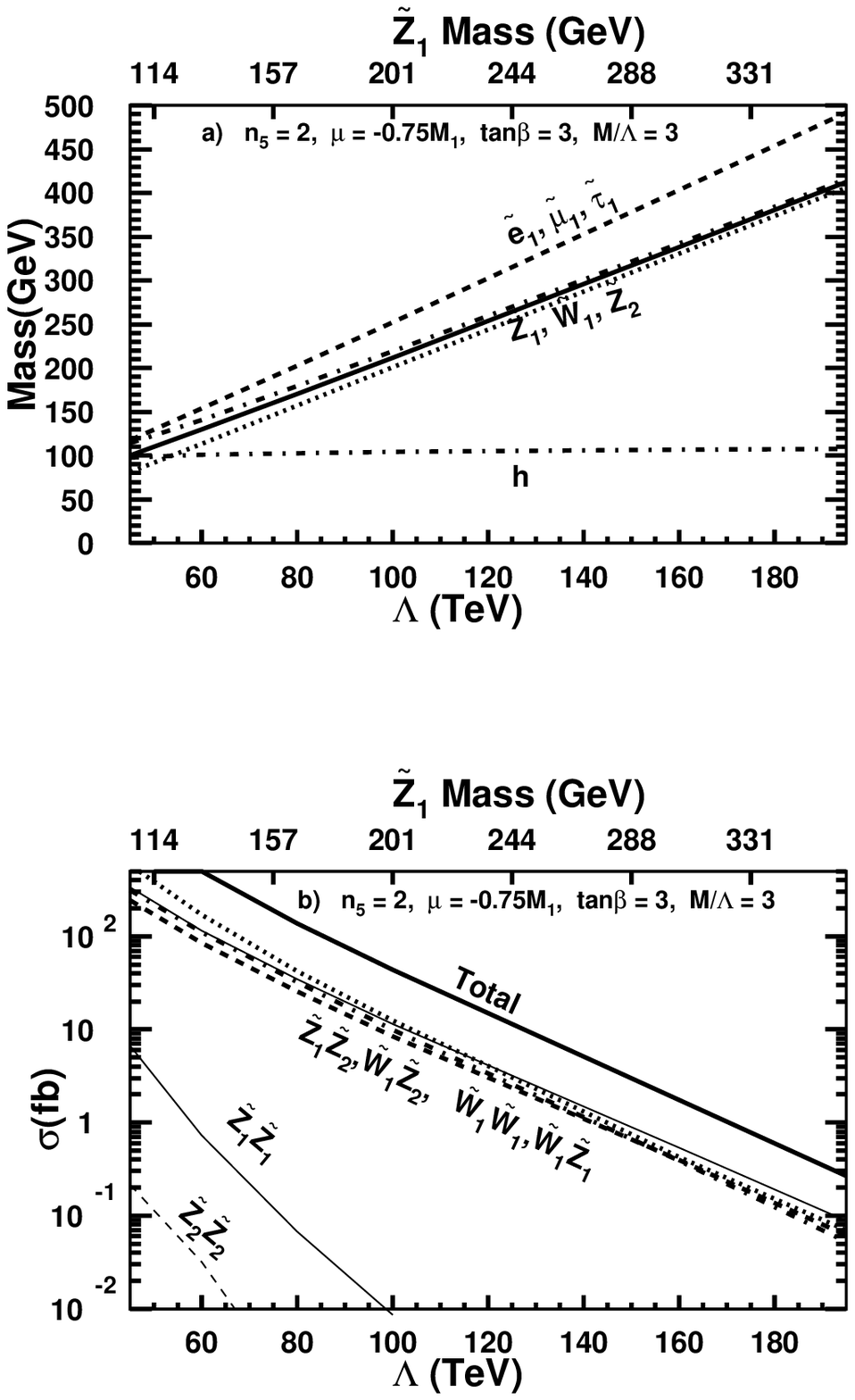}
\caption[]{The same as Fig.~\ref{mlA:prod}, but for the Higgsino NLSP
model line D. In frame {\it a}) the dotted line denotes the lightest
neutralino, the solid line denotes the lighter chargino and the upper
dot-dashed line denotes the second lightest neutralino. In frame {\it
b}) the dashed, dot-dashed, solid and dotted lines denote  cross
sections for $\tz_1
\tz_2$, $\tw_1 \tz_2$, $\tw_1 \tw_1$, $\tw_1 \tz_1$ production,
respectively.}
\label{mlD:prod}
\end{figure}

%Fig. 14
\begin{figure}
\dofig{6in}{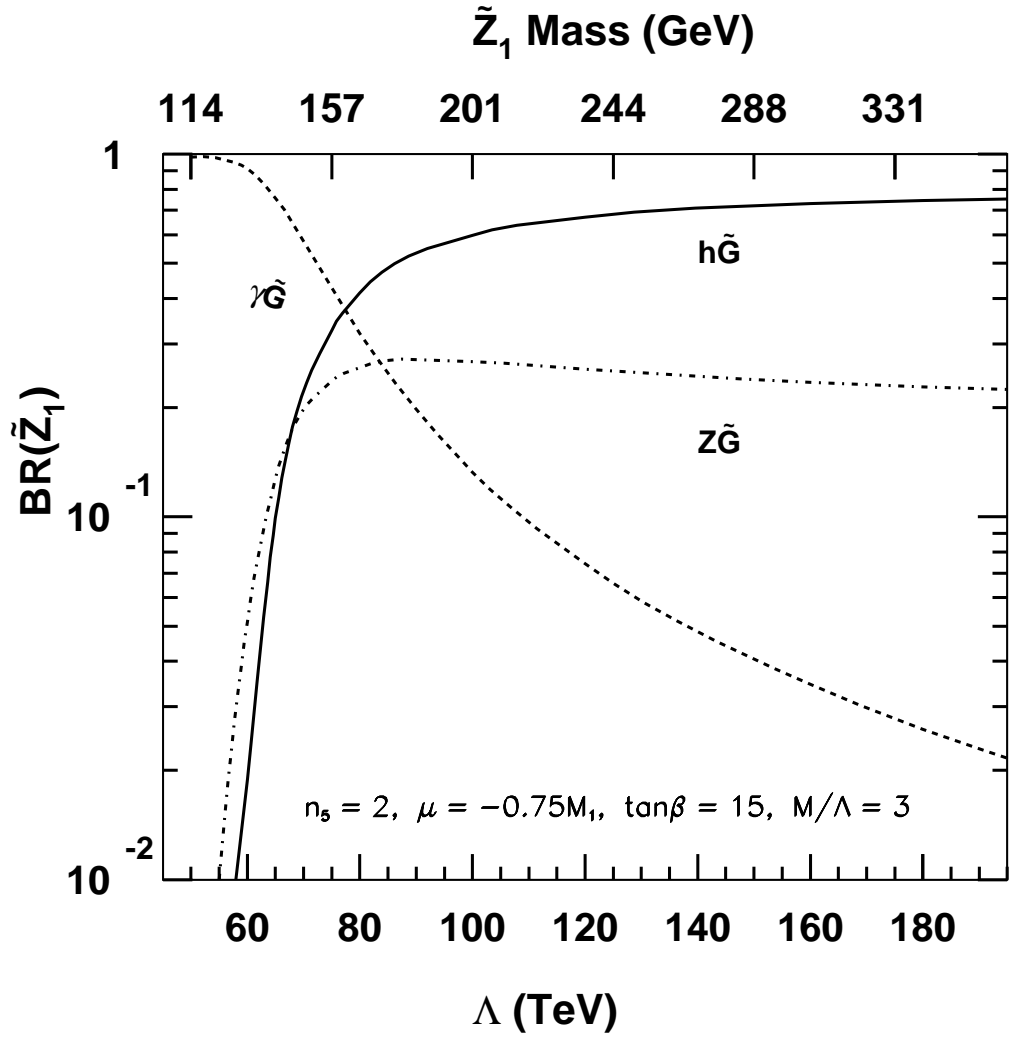}
\caption[]{Branching ratios for various decays of the neutralino NLSP
versus $\Lambda$ for model line D. The upper scale shows the neutralino
mass.}
\label{mlD:decay}
\end{figure}

%Fig. 15
\begin{figure}
\dofig{6in}{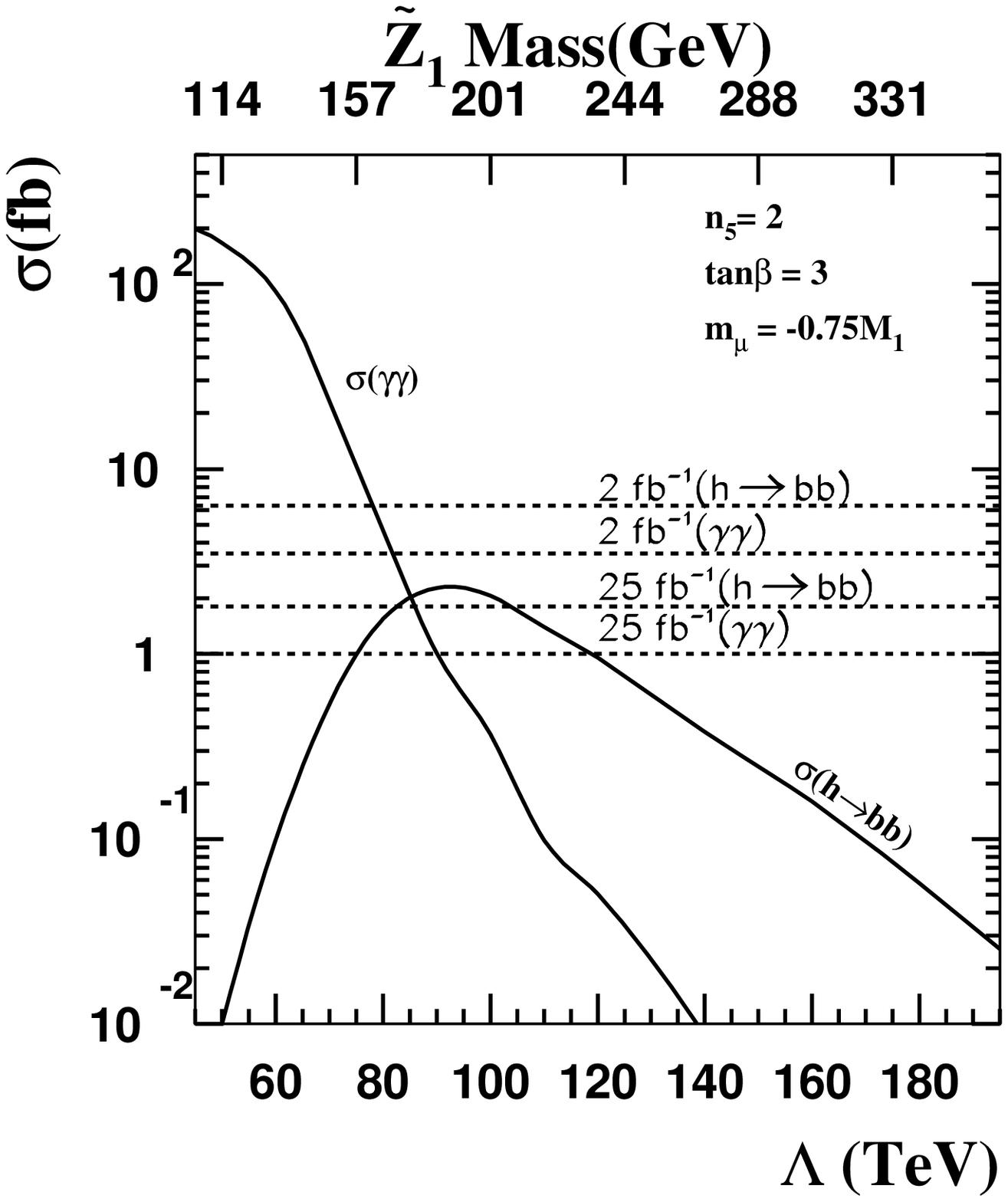}
\caption[]{SUSY signal cross sections for the inclusive $\gamma\gamma +
\eslt$ events (labelled $\sigma(\gamma\gamma)$, and for events with $\geq
3$ tagged $b$-jets (labelled $\sigma(h \to bb)$ after all cuts described
in the text for model line~D. The dashed horizontal lines denote the
minimum cross section for the signal to be observable at the $5\sigma$
level at the MI and at TeV33.}
\label{mlD:reach}
\end{figure}


\begin{references}
\bibitem{footnote3} In addition to these decays, ISAJET also includes several
other three body decays of $\tell_2 \simeq \tell_L,\ttau_2$ and $\tnu_{\ell},
\tnu_{\tau}$ in various approximations. These decays are unimportant for us
because the heavier sleptons generally have two-body
decays, but might be important in models~\cite{murayama} where these are
light. The reader who uses ISAJET in models where these decays
are important is advised to become aware of the various approximations under
which these are calculated.
\bibitem{footnote5} Although the chargino pair production cross
section is an order of magnitude larger than the slepton cross section,
it should be remembered that the branching fraction for both charginos
to decay to $e$ or to $\mu$ is just $\sim 2$\%.
\bibitem{footnote6} Admittedly, we have not attempted to separate
slepton events from other SUSY sources. These additional SUSY sources
dominate for $\Lambda \leq 90$~GeV for which $\tz_2 \to \ell \tell_1$ is
kinematically accessible and has a large branching ratio. Lepton pairs
from neutralino decays will necessarily have $m(\ell\ell) \leq
m_{\tz_2}\sqrt{1-\frac{m_{\tell_1}^2}{m_{\tz_2}^2}}\sqrt{1-\frac{m_{\tz_1}^2}{m_{\tell_1}^2}}$,
while those from slepton pair production form a continuum, possibly
extending beyond this.
\bibitem{footnote7} We have checked that this additional
requirement causes essentially no loss of signal.
\bibitem{footnote8} We studied several other distributions including, jet
multiplicity, $\theta_{bb}$, $\delta\phi(bb)$ but none of these proved
useful to enhance the signal over the top background.
\end{references}
\end{document}